\documentclass[iop,apj]{emulateapj}
\usepackage{graphicx,url}

\shorttitle{Late-Forming dIrrs}
\shortauthors{Kirby et al.}
\begin{document}

\newcommand{\nmem}{197}
\newcommand{\nnonmem}{123}
\newcommand{\nfeh}{179}
\newcommand{\leoan}{127}
\newcommand{\leoafehmean}{-1.67}
\newcommand{\leoafehmeanerr}{0.08}
\newcommand{\leoafehmeanerrlower}{0.08}
\newcommand{\leoafehmeanerrupper}{0.09}
\newcommand{\leoafehsigma}{0.39}
\newcommand{\leoafehsigmaerr}{0.06}
\newcommand{\leoafehsigmaerrlower}{0.06}
\newcommand{\leoafehsigmaerrupper}{0.07}
\newcommand{\leoags}{2.3}
\newcommand{\leoagserr}{0.5}
\newcommand{\leoameanv}{26.2}
\newcommand{\leoameanverr}{0.9}
\newcommand{\leoameanverrlower}{0.9}
\newcommand{\leoameanverrupper}{1.0}
\newcommand{\leoasigmav}{ 9.0}
\newcommand{\leoasigmaverr}{ 0.7}
\newcommand{\leoasigmaverrlower}{ 0.6}
\newcommand{\leoasigmaverrupper}{ 0.8}
\newcommand{\leoaml}{12}
\newcommand{\leoamlerrlower}{ 3}
\newcommand{\leoamlerrupper}{ 3}
\newcommand{\leoamm}{7.3}
\newcommand{\leoammerrlower}{1.0}
\newcommand{\leoammerrupper}{1.1}
\newcommand{\leoavgsr}{ -13.9}
\newcommand{\leoadensity}{0.1}
\newcommand{\leoadensityerrl}{0.0}
\newcommand{\leoadensityerru}{ 0.0}
\newcommand{\aqrn}{25}
\newcommand{\aqrfehmean}{-1.50}
\newcommand{\aqrfehmeanerr}{0.06}
\newcommand{\aqrfehmeanerrlower}{0.06}
\newcommand{\aqrfehmeanerrupper}{0.06}
\newcommand{\aqrfehsigma}{0.10}
\newcommand{\aqrfehsigmaerr}{0.07}
\newcommand{\aqrfehsigmaerrlower}{0.06}
\newcommand{\aqrfehsigmaerrupper}{0.09}
\newcommand{\aqrgs}{1.4}
\newcommand{\aqrgserr}{0.2}
\newcommand{\aqrmeanv}{-141.8}
\newcommand{\aqrmeanverr}{1.9}
\newcommand{\aqrmeanverrlower}{2.0}
\newcommand{\aqrmeanverrupper}{1.8}
\newcommand{\aqrsigmav}{ 7.8}
\newcommand{\aqrsigmaverr}{ 1.5}
\newcommand{\aqrsigmaverrlower}{ 1.1}
\newcommand{\aqrsigmaverrupper}{ 1.8}
\newcommand{\aqrml}{21}
\newcommand{\aqrmlerrlower}{ 4}
\newcommand{\aqrmlerrupper}{ 6}
\newcommand{\aqrmm}{9.6}
\newcommand{\aqrmmerrlower}{1.8}
\newcommand{\aqrmmerrupper}{2.5}
\newcommand{\aqrvgsr}{ -30.7}
\newcommand{\aqrdensity}{0.1}
\newcommand{\aqrdensityerrl}{0.0}
\newcommand{\aqrdensityerru}{ 0.1}
\newcommand{\sagdign}{45}
\newcommand{\sagdigfehmean}{-1.88}
\newcommand{\sagdigfehmeanerr}{0.11}
\newcommand{\sagdigfehmeanerrlower}{0.09}
\newcommand{\sagdigfehmeanerrupper}{0.13}
\newcommand{\sagdigfehsigma}{0.28}
\newcommand{\sagdigfehsigmaerr}{0.09}
\newcommand{\sagdigfehsigmaerrlower}{0.03}
\newcommand{\sagdigfehsigmaerrupper}{0.15}
\newcommand{\sagdiggs}{4.6}
\newcommand{\sagdiggserr}{1.2}
\newcommand{\sagdigmeanv}{-78.4}
\newcommand{\sagdigmeanverr}{1.6}
\newcommand{\sagdigmeanverrlower}{1.6}
\newcommand{\sagdigmeanverrupper}{1.6}
\newcommand{\sagdigsigmav}{ 9.4}
\newcommand{\sagdigsigmaverr}{ 1.3}
\newcommand{\sagdigsigmaverrlower}{ 1.1}
\newcommand{\sagdigsigmaverrupper}{ 1.5}
\newcommand{\sagdigml}{10}
\newcommand{\sagdigmlerrlower}{ 3}
\newcommand{\sagdigmlerrupper}{ 3}
\newcommand{\sagdigmm}{4.4}
\newcommand{\sagdigmmerrlower}{0.8}
\newcommand{\sagdigmmerrupper}{1.0}
\newcommand{\sagdigvgsr}{   6.2}
\newcommand{\sagdigdensity}{0.3}
\newcommand{\sagdigdensityerrl}{0.1}
\newcommand{\sagdigdensityerru}{ 0.1}
\newcommand{\sagdigoverlap}{30}
\newcommand{\sagdigmommem}{22}
\newcommand{\sagdigmomnm}{8}
\newcommand{\sagdigpmmem}{22}
\newcommand{\sagdigpmnm}{1}
\newcommand{\leoavrotlim}{5.0}
\newcommand{\leoavsigmalim}{0.6}
\newcommand{\aqrvrotlim}{9.0}
\newcommand{\aqrvsigmalim}{1.1}
\newcommand{\sagdigvrotlim}{6.0}
\newcommand{\sagdigvsigmalim}{0.6}
\newcommand{\leoahivmean}{23.7}
\newcommand{\leoahisigmavmean}{6.2}
\newcommand{\aqrhivmean}{-140.3}
\newcommand{\aqrhisigmavmean}{6.7}
\newcommand{\sagdighivmean}{-79.2}
\newcommand{\sagdighisigmavmean}{8.2}
\newcommand{\leoaagefehint}{-1.42}
\newcommand{\leoaagefehinterr}{0.07}
\newcommand{\leoaagefehslope}{0.03}
\newcommand{\leoaagefehslopeerr}{0.01}
\newcommand{\aqragefehint}{-0.67}
\newcommand{\aqragefehinterr}{0.37}
\newcommand{\aqragefehslope}{0.30}
\newcommand{\aqragefehslopeerr}{0.13}
\newcommand{\sagdigagefehint}{-1.86}
\newcommand{\sagdigagefehinterr}{0.13}
\newcommand{\sagdigagefehslope}{0.00}
\newcommand{\sagdigagefehslopeerr}{0.00}
\newcommand{\leoafehkurtosis}{0.12}
\newcommand{\leoafehkurtosiserr}{0.45}
\newcommand{\leoafehskewness}{0.10}
\newcommand{\leoafehskewnesserr}{0.23}
\newcommand{\leoafehinitial}{-2.47}
\newcommand{\leoafehinitialerrl}{+0.12}
\newcommand{\leoafehinitialerru}{+0.10}
\newcommand{\leoayieldsimple}{0.035}
\newcommand{\leoayieldsimpleerru}{0.004}
\newcommand{\leoayieldsimpleerrl}{0.004}
\newcommand{\leoayieldpre}{0.026}
\newcommand{\leoayieldpreerrl}{0.003}
\newcommand{\leoayieldpreerru}{0.004}
\newcommand{\leoainfallm}{6.0}
\newcommand{\leoainfallmerrl}{2.0}
\newcommand{\leoainfallmerru}{3.2}
\newcommand{\leoayieldinfall}{0.032}
\newcommand{\leoayieldinfallerrl}{0.003}
\newcommand{\leoayieldinfallerru}{0.003}
\newcommand{\leoaaiccsimple}{121.21}
\newcommand{\leoaaiccpre}{96.94}
\newcommand{\leoaaiccinfall}{99.15}
\newcommand{\leoaaiccprediff}{24.27}
\newcommand{\leoaaiccinfalldiff}{22.06}
\newcommand{\leoappreinfall}{0.33}
\newcommand{\aqrfehkurtosis}{-0.77}
\newcommand{\aqrfehkurtosiserr}{0.93}
\newcommand{\aqrfehskewness}{0.35}
\newcommand{\aqrfehskewnesserr}{0.48}
\newcommand{\aqrfehinitial}{*****}
\newcommand{\aqrfehinitialerrl}{+7.86}
\newcommand{\aqrfehinitialerru}{*****}
\newcommand{\aqryieldsimple}{0.055}
\newcommand{\aqryieldsimpleerru}{0.012}
\newcommand{\aqryieldsimpleerrl}{0.011}
\newcommand{\aqryieldpre}{0.053}
\newcommand{\aqryieldpreerrl}{0.012}
\newcommand{\aqryieldpreerru}{0.014}
\newcommand{\aqrinfallm}{7.1}
\newcommand{\aqrinfallmerrl}{3.8}
\newcommand{\aqrinfallmerru}{6.2}
\newcommand{\aqryieldinfall}{0.044}
\newcommand{\aqryieldinfallerrl}{0.007}
\newcommand{\aqryieldinfallerru}{0.008}
\newcommand{\aqraiccsimple}{28.19}
\newcommand{\aqraiccpre}{31.55}
\newcommand{\aqraiccinfall}{27.45}
\newcommand{\aqraiccprediff}{-3.36}
\newcommand{\aqraiccinfalldiff}{0.74}
\newcommand{\aqrppreinfall}{7.77}
\newcommand{\sagdigfehkurtosis}{-0.03}
\newcommand{\sagdigfehkurtosiserr}{0.71}
\newcommand{\sagdigfehskewness}{0.24}
\newcommand{\sagdigfehskewnesserr}{0.36}
\newcommand{\sagdigfehinitial}{*****}
\newcommand{\sagdigfehinitialerrl}{+5.03}
\newcommand{\sagdigfehinitialerru}{*****}
\newcommand{\sagdigyieldsimple}{0.024}
\newcommand{\sagdigyieldsimpleerru}{0.005}
\newcommand{\sagdigyieldsimpleerrl}{0.004}
\newcommand{\sagdigyieldpre}{0.024}
\newcommand{\sagdigyieldpreerrl}{0.004}
\newcommand{\sagdigyieldpreerru}{0.005}
\newcommand{\sagdiginfallm}{2.3}
\newcommand{\sagdiginfallmerrl}{0.9}
\newcommand{\sagdiginfallmerru}{1.9}
\newcommand{\sagdigyieldinfall}{0.021}
\newcommand{\sagdigyieldinfallerrl}{0.003}
\newcommand{\sagdigyieldinfallerru}{0.004}
\newcommand{\sagdigaiccsimple}{65.89}
\newcommand{\sagdigaiccpre}{69.32}
\newcommand{\sagdigaiccinfall}{68.28}
\newcommand{\sagdigaiccprediff}{-3.44}
\newcommand{\sagdigaiccinfalldiff}{-2.39}
\newcommand{\sagdigppreinfall}{1.69}
\newcommand{\mathfeh}{{\rm [Fe/H]}}

\title{Chemistry and Kinematics of the Late-Forming Dwarf Irregular
  Galaxies Leo~A, Aquarius, and Sagittarius DIG\altaffilmark{*}}

\author{Evan~N.~Kirby\altaffilmark{1},
  Luca~Rizzi\altaffilmark{2},
  Enrico~V.~Held\altaffilmark{3},
  Judith~G.~Cohen\altaffilmark{1},
  Andrew~A.~Cole\altaffilmark{4},
  Ellen~M.~Manning\altaffilmark{4},
  Evan~D.~Skillman\altaffilmark{5},
  Daniel~R.~Weisz\altaffilmark{6}}

\altaffiltext{*}{The data presented herein were obtained at the
  W.~M.~Keck Observatory, which is operated as a scientific
  partnership among the California Institute of Technology, the
  University of California and the National Aeronautics and Space
  Administration. The Observatory was made possible by the generous
  financial support of the W.~M.~Keck Foundation.}
\altaffiltext{1}{California Institute of Technology, 1200 E.\ California Blvd., MC 249-17, Pasadena, CA 91125, USA}
\altaffiltext{2}{W.M.\ Keck Observatory, 65-1120 Mamalahoa Hwy., Kamuela, HI 96743, USA}
\altaffiltext{3}{INAF, Osservatorio Astronomico di Padova, Vicolo dell'Osservatorio 5, 35122, Padova, Italy}
\altaffiltext{4}{School of Physical Sciences, University of Tasmania, Private Bag 37, Hobart, Tasmania, 7001 Australia}
\altaffiltext{5}{Minnesota Institute for Astrophysics, University of Minnesota, Minneapolis, MN 55455, USA}
\altaffiltext{6}{University of California, Berkeley, CA 94720, USA}



\begin{abstract}

We present Keck/DEIMOS spectroscopy of individual stars in the
relatively isolated Local Group dwarf galaxies Leo~A, Aquarius, and
the Sagittarius dwarf irregular galaxy.  The three galaxies---but
especially Leo~A and Aquarius---share in common delayed star formation
histories relative to many other isolated dwarf galaxies.  The stars
in all three galaxies are supported by dispersion.  We found no
evidence of stellar velocity structure, even for Aquarius, which has
rotating H$\,${\sc i} gas.  The velocity dispersions indicate that all
three galaxies are dark matter-dominated, with dark-to-baryonic mass
ratios ranging from
$\sagdigmm_{-\sagdigmmerrlower}^{+\sagdigmmerrupper}$ (SagDIG) to
$\aqrmm_{-\aqrmmerrlower}^{+\aqrmmerrupper}$ (Aquarius).  Leo~A and
SagDIG have lower stellar metallicities than Aquarius, and they also
have higher gas fractions, both of which would be expected if Aquarius
were farther along in its chemical evolution.  The metallicity
distribution of Leo~A is inconsistent with a Closed or Leaky Box model
of chemical evolution, suggesting that the galaxy was pre-enriched or
acquired external gas during star formation.  The metallicities of
stars increased steadily for all three galaxies, but possibly at
different rates.  The [$\alpha$/Fe] ratios at a given [Fe/H] are lower
than that of the Sculptor dwarf spheroidal galaxy, which indicates
more extended star formation histories than Sculptor, consistent with
photometrically derived star formation histories.  Overall, the bulk
kinematic and chemical properties for the late-forming dwarf galaxies
do not diverge significantly from those of less delayed dwarf
galaxies, including dwarf spheroidal galaxies.

\end{abstract}


\section{Introduction}
\label{sec:intro}

The Local Group plays host to dozens of dwarf galaxies.  These
galaxies are laboratories for star formation and chemical evolution
because they span a huge range of age, stellar mass, metallicity,
morphology, and gas fraction.  A glance at \textit{Hubble Space
  Telescope} (HST) color-magnitude diagrams of Local Group dwarf
galaxies reveals an expansive diversity of stellar populations
\citep{hol06,wei14a}.  For example, deep HST imaging of the dwarf
spheroidal galaxy (dSph) Cetus shows no evidence of young stars
\citep{mon10b}, whereas similarly deep HST imaging of the dwarf
irregular galaxy (dIrr) Aquarius shows old red giants, an
intermediate-age main sequence turn-off, and young stars on the main
sequence \citep{col14}.  Both Cetus and Aquarius are isolated and have
stellar masses around $2 \times 10^6~M_{\sun}$, yet their star
formation histories (SFHs) are very different.

It is clear that environment plays a big role in shaping the SFHs and
the gas content of these galaxies \citep[e.g.,][]{van94}.  For
example, no dwarf galaxy within $\sim 300$~kpc of the Milky Way (MW)
except the Magellanic Clouds has any gas \citep{grc09,spe14}.
Furthermore, all of these satellite galaxies are old, spheroidal, and
supported by dispersion rather than rotation.  On the other hand, with
only two exceptions \citep[Cetus and Tucana,][]{mon10b,mon10a}, all of
the Local Group dwarf galaxies farther than 300~kpc have gas and at
least some young stars.  They also tend to be characterized by
irregularly distributed young stellar populations, and some of them
rotate.  The environment effect persists far outside the Local Group.
Nearly all galaxies with $M_* > 10^7~M_{\sun}$ farther than 1.5~Mpc
from a massive host galaxy are presently forming stars, but proximity
to a host can terminate star formation \citep{geh12}.  However, it is
worth noting that present isolation does not ensure that a galaxy was
always isolated.  For example, one possible reason for Cetus and
Tucana's lack of star formation is that they were once close enough to
the Milky Way or M31 to experience ram pressure and/or tidal stripping
\citep{tey12}.

While it is clear that many dwarf galaxies lack young populations,
every dwarf galaxy contains at least some ancient stars older than
10~Gyr \citep{mat98,orb08,wei14a}.  Both isolated and satellite
galaxies can have intermediate-aged stellar populations.  Examples
include the satellite dSphs Fornax and Leo~I and the relatively
isolated dIrrs IC~1613 and NGC~6822.  Apparently, all dwarf galaxies
start with at least a skeleton population of ancient stars.  Some
galaxies manage to build later populations whereas others---almost
entirely satellites of larger hosts---do not.

However, the stellar chemical properties of the dwarf galaxies do not
depend on environment to nearly the same degree as the gas content.
All dwarf galaxies obey the same relationship between stellar mass and
stellar metallicity, regardless of environment or SFH
\citep{ski89a,kir13b}.  Therefore, metallicity and chemical evolution
seem to be inextricably tied to the stellar mass of a galaxy.
Although environment influences gas fraction and SFH, it does not
overtly influence chemistry.

The galaxies with the best-constrained star formation histories are
those with space-based photometry that reaches at least as deep as the
main sequence turn-off (MSTO) of the oldest stellar population.  Such
data is especially resource-intensive for the isolated galaxies, which
are farther away than MW satellite galaxies.  The Local Cosmology from
Isolated Dwarfs \citep[LCID,][]{ber08,ber09} project has collected HST
imaging of isolated dwarfs.  \citet{ski14} compared the SFHs of the
six LCID dwarfs: the dSphs Cetus and Tucana, the dIrrs IC~1613 and
Leo~A, and the transition dwarfs LGS~3 and Phoenix.  To date, the only
other isolated galaxies with comparable data quality are Leo~T
\citep{wei12} and Aquarius \citep{col14}.

With this deep HST imaging, \citet{col07,col14} found that Leo~A
(DDO~69) and Aquarius (DDO~210) both experienced highly delayed star
formation relative to other dIrrs.  Like all dwarf galaxies, Leo~A and
Aquarius have ancient populations.  However, they distinguish
themselves by having formed less than 20\% of their stars more than
6--8~Gyr ago.  They are also notable for being among the most isolated
galaxies in the local Galactic neighborhood.  Both galaxies are
probably members of the Local Group.  Although they lie right at the
zero-velocity surface that separates the Local Group from the Hubble
flow \citep{kar09}, their velocities relative to the Local Group
barycenter suggest that they are bound to the group \citep{mcc12}.
However, their free-fall times to the MW, M31, or the Local Group
barycenter are on the order of a Hubble time, which means that they
have been living and will continue to live in the most remote regions
of the Local Group essentially forever.

The isolation and delayed SFHs of Leo~A and Aquarius led us to conduct
a spectroscopic survey of their red giants.  We also included the
Sagittarius dIrr (SagDIG) in this study because it is also as isolated
as the other two dIrrs.  The gas and stellar populations of all three
dIrrs have low probabilities ($\le 2\%$) of ever having been
influenced by the Milky Way \citep{tey12}.  SagDIG's SFH is not as
extreme as Leo~A or Aquarius \citep{wei14a}.  Instead, about 40\% of
its stars are ancient.  However, like Leo~A and Aquarius, it then
stopped forming stars for several Gyr before restarting.  It is worth
keeping in mind that the interpretation of SagDIG's SFH is based on
shallower HST imaging that does not reach the ancient MSTO\@.  We now
discuss some other interesting properties of each galaxy to keep in
mind when interpreting the spectroscopic data.

\subsection{Leo~A}
\label{sec:leoa}

Using HST/WFPC2 photometry, \citet{tol98} first noticed that Leo~A
conspicuously lacked a prominent ancient ($>10$~Gyr) population.  For
that reason, it stood out among the Local Group and challenged the
idea that all dwarf galaxies are old.  The claim of the absence of an
old population was not without controversy.  Also using HST/WFPC2
photometry, \citet{sch02} claimed that Leo~A was predominantly
ancient, and \citet{dol02,dol03} confirmed the presence of a small
number of RR~Lyrae, which are necessarily ancient, as well as an old,
metal-poor halo around the galaxy.  It was not until \citet{col07}
obtained HST imaging down to the old MSTO that the fraction of ancient
stars was conclusively determined to be less than 10\%.

The SFH that \citeauthor{col07}\ derived shows that most of the star
formation happened more recently than 6~Gyr ago, and it ramped up from
then until 2~Gyr ago.  The present star formation rate (SFR) is
measured as $9.3 \times 10^{-5}~M_{\sun}$~yr$^{-1}$ from H$\alpha$
imaging and $6.0 \times 10^{-4}~M_{\sun}$~yr$^{-1}$ from ultraviolet
(UV) imaging \citep{kar13}.  The discrepancy for dwarf galaxies is
well-known \citep{lee09}, but the UV rate is generally regarded as
more robust.  At the present UV-measured SFR, Leo~A could have formed
all of its stars in 5.5~Gyr, which is broadly consistent with the late
SFH observed by \citet{col07}.

The best constraints from broadband photometry have not detected
long-term trends in the average stellar metallicity, which appears to
be constant at $\rm{[Fe/H]} = -1.4$ \citep{col07}.  A peculiarly low
H$\,${\sc ii} region metallicity \citep[${\rm [O/H]} = -1.31 \pm
  0.10$,][]{vanzee06} and consequently low effective yield
\citep{lee06} also make Leo~A stand out among dIrrs.  The previously
measured average stellar metallicity \citep[$\langle {\rm [Fe/H]}
  \rangle = -1.58$,][]{kir13b} is not far below the gas-phase
metallicity.  The potential similarity in metallicity between stars
and gas possibly corroborates the slow evolution in metallicity
observed by \citet{col07}.  In other words, the present (gas-phase)
metallicity has not progressed much beyond the past-averaged (stellar)
metallicity.

Early H$\,${\sc i} maps hinted that Leo~A is not rotationally
supported \citep{all78}.  Higher resolution maps showed some rotation
of the gas, but the rotation is not aligned with the stellar disk
\citep{you96}.  A large, low-velocity feature near the center of the
galaxy---near the bulk of the stars---could be seen as the receding
half of a nearly edge-on disk, but \citeauthor{you96}\ suggested it is
a contracting or expanding shell, possibly due to a recent supernova.
Instead of rotation, they measured a mostly dispersion-supported
galaxy with $\sigma_v = 9$~km~s$^{-1}$.  This dispersion could have
been consistent with no dark matter if the H$\,${\sc i} gas is
supported by its own internal pressure.  However, \citet{bro07}
measured the stellar velocity dispersion from 10 B supergiants.  They
also obtained $\sigma_v = 9$~km~s$^{-1}$.  Later, \citet{kir14}
measured $\sigma_v = 6.4_{-1.2}^{+1.4}$~km~s$^{-1}$ from red giants.
Either measurement of the stellar velocity dispersion conclusively
indicates that the galaxy is supported by the dynamical pressure
provided by a dark matter subhalo rather than hydrodynamical pressure.

\subsection{Aquarius}
\label{sec:aqr}

Other than the HST-based SFH of \citet{col14}, Aquarius is the least
well studied of the three galaxies presented here.  A few surveys have
examined the carbon stars \citep{gul07} and variable stars
\citep{ord16} in Aquarius.  The latter study found both RR~Lyrae and
Cepheids, indicating the presence of ancient ($>10$~Gyr) and fairly
young ($\sim 300$~Myr) populations.  Interestingly, the young Cepheids
were offset from the galactic center, where the older Cepheids were
found.  \citeauthor{ord16}\ interpreted this displacement as stellar
migration, though the stars could have formed in an off-center star
formation region.

Although \citet{col14} found that star formation in Aquarius
experienced a late resurgence, H$\alpha$ and ultraviolet imaging show
that its current SFR is zero \citep{van00,hun10}.  In contrast to
Leo~A, Aquarius seems to have tapered off its SFR recently.  It isn't
clear whether the current lack of star formation is permanent or a
temporary lull, possibly as the result of recent stellar feedback.
The presence of copious H$\,${\sc i} gas \citep{lo93} suggests that
the galaxy is not yet finished with star formation.

\subsection{SagDIG}
\label{sec:sagdig}

SagDIG has the highest gas fraction ($M$(H$\,${\sc i})$/M_* =
\sagdiggs \pm \sagdiggserr$) of any galaxy in the Local Group.  The
high gas fraction also imparts a high specific SFR to SagDIG\@.  The
UV-measured SFR is $7.2 \times 10^{-4}~M_{\sun}~{\rm yr}^{-1}$
\citep{kar13}, corresponding to a specific SFR of $0.4~{\rm
  Gyr}^{-1}$.  That makes SagDIG one of the fastest growing galaxies
in the Local Group.  It could have grown to its present stellar mass
at its present SFR in just 2.5~Gyr.

SagDIG's high specific SFR makes its metallicity very interesting.  It
has a very low gas-phase metallicity \citep[${\rm [O/H]} =
  -1.4$,][]{ski89b,sav02}.  That ties SagDIG with Leo~A for the
distinction of being the most oxygen-poor galaxy in the Local Group,
with an oxygen abundance nearly as low as the quintessentially
metal-poor galaxy I~Zw~18 \citep{sea72,ski93}.

Ground-based, wide-field imaging shows that the stellar population of
SagDIG is elongated and very extended.  The surface brightness profile
drops exponentially out to 5~arcmin, at which point it fades into the
background \citep{bec14}.  The galaxy may even be embedded in a very
low-density stellar halo \citep{hig16}.  Curiously, the stellar
distribution seems to have little to do with the H$\,${\sc i}
distribution.  Whereas the stellar distribution is smooth and
elongated, the neutral gas is round.  It is also asymmetric, with an
apparent hole just to the southwest of the galaxy's center.

A Closed Box model of chemical evolution coupled with the measured gas
fraction (revised for possible gas loss) and gas-phase metallicity
suggests that the average stellar metallicity of SagDIG should be
${\rm [Fe/H]} = -2$ \citep{sav02}.  This value is exactly in accord
with photometric estimates of the stellar metallicity
\citep{mom02,mom05}.  However, it is $2.3\sigma$ lower than the
stellar metallicity predicted by the spectroscopically measured
stellar mass--stellar metallicity relation \citep{kir13b}.  A
spectroscopic measurement of the stellar metallicity would answer
whether the Closed Box model and the photometric estimate of the
metallicity are valid or whether SagDIG conforms to the
mass--metallicity relation.


\section{Spectroscopic Observations}
\label{sec:obs}

\citet{kir14} already published some Keck/DEIMOS \citep{fab03}
spectroscopy of stars in Leo~A and Aquarius.  We obtained additional
DEIMOS spectra of individual stars in those galaxies as well as
SagDIG\@.  This section describes those observations.

\subsection{Source Catalogs}

We designed DEIMOS slitmasks using photometry and astrometry from
multiple sources.  For Leo~A, we used the $V$ and $I$
Subaru/Suprime-Cam \citep{miy02} catalog of \citet{sto14}.  For
Aquarius, we used the catalog of \citet{mcc06}, who observed the
galaxy with Suprime-Cam in the $V$ and $I$ filters.  For SagDIG, we
used two different photometry catalogs.  We combined
\citeauthor{mom02}'s (\citeyear{mom02}) photometry from the ESO
Multimode Instrument \citep{dek86} at the ESO New Technology Telescope
(NTT) with \citeauthor{mom14}'s (\citeyear{mom14}) photometry from the
Hubble Space Telescope/Advanced Camera for Surveys (ACS).  Although
the ACS photometry is more accurate than the NTT photometry, the field
of view of ACS is only 3.4$\arcmin$ square.  Hence, we supplemented
the ACS photometry with the NTT photometry, which spans a 6.2$\arcmin$
square.  The NTT photometry was obtained in $V$ and $I$ filters, and
we converted the ACS F475W, F606W, and F814W magnitudes into $V$ and
$I$ with the transformation equations of \citet{sah11}.  All
magnitudes were corrected for extinction with the \citet{sch98} dust
maps.

\subsection{Target Selection}
\label{sec:target}

\begin{figure*}[t!]
\centering
\includegraphics[width=0.8\textwidth]{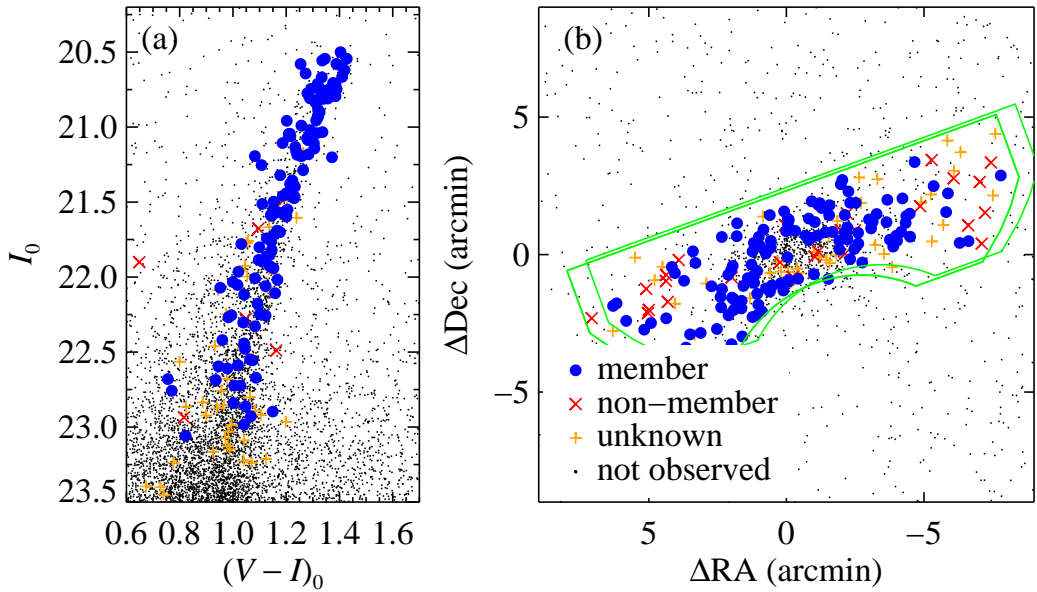}
\caption{(a) The color--magnitude diagram of Leo~A from
  \citeauthor{sto14}'s (\citeyear{sto14}) photometry catalog.  Stars
  confirmed to be members are shown as blue points.  Non-members are
  shown as red crosses; stars of inconclusive membership are shown as
  orange plusses; and objects not observed with DEIMOS are shown as
  black points.  (b) The sky in the area of the DEIMOS observations.
  The outlines of the DEIMOS slitmasks are shown in
  green.\label{fig:leoa}}
\end{figure*}

\begin{figure*}[t!]
\centering
\includegraphics[width=0.8\textwidth]{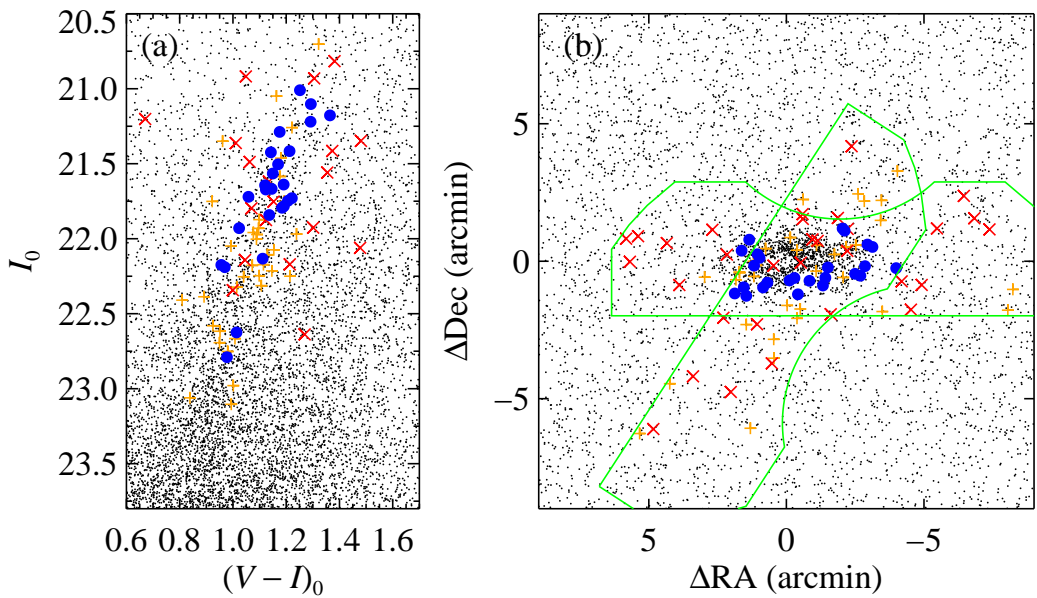}
\caption{Same as Figure~\ref{fig:leoa} but for Aquarius.  The
  photometry comes from \citet{mcc06}.\label{fig:aqr}}
\end{figure*}

\begin{figure*}[t!]
\centering
\includegraphics[width=0.8\textwidth]{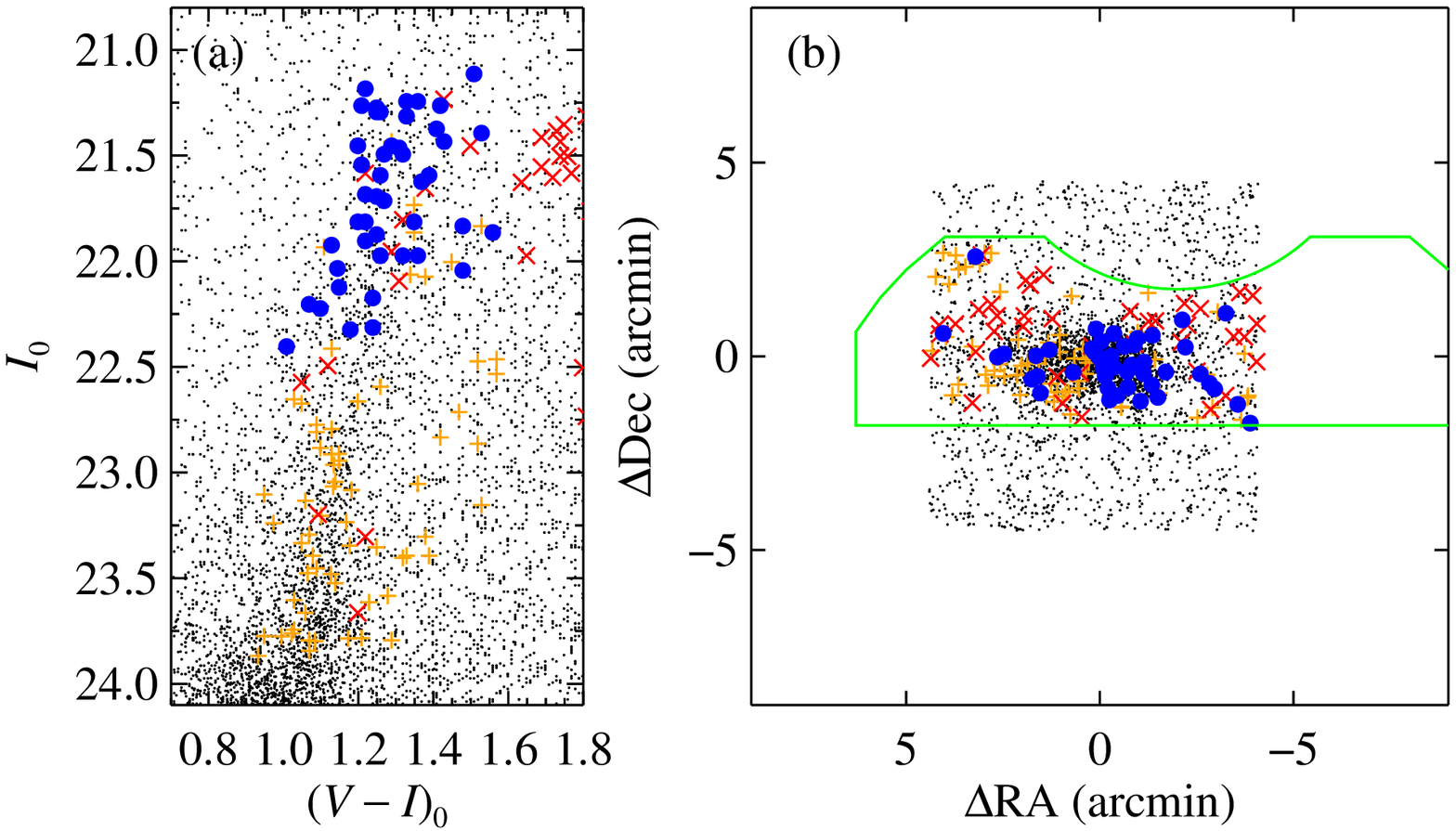}
\caption{Same as Figure~\ref{fig:leoa} but for SagDIG\@.  The
  photometry comes from \citet{mom02,mom14}.\label{fig:sagdig}}
\end{figure*}

DEIMOS slitmask design constraints required that we prioritize targets
for placement on the slitmasks.  We first prioritized stars that were
likely to be members of the red giant branch (RGB) of their respective
galaxy.  We identified the RGB by overlaying theoretical isochrones on
the color--magnitude diagram (CMD).  We used Yonsei--Yale isochrones
in the $V$ and $I$ filters \citep{dem04}.  The isochrones were shifted
to the distance modulus appropriate for each galaxy: $(m-M)_0 = 24.59$
for Leo~A \citep{dol02,tam11}, 24.95 for Aquarius \citep{col14}, and
25.10 for SagDIG \citep{mom05}

We chose two isochrones to bound the red and blue sides of the RGB\@.
The red side had an age of 14~Gyr and a metallicity of $\rm{[Fe/H]} =
0$.  The blue side had an age of 2~Gyr and the minimum metallicity of
the isochrone set: $\rm{[Fe/H]} = -3.8$ and $-2.2$ for Yonsei--Yale
and Padova, respectively.  Stars near the middle of the color range of
these two bounding isochrones were given higher priority.  Brighter
stars were also given higher priority.  Finally, we selected stars
outside of the red and blue bounding isochrones to fill any remaining
space on the slitmask.  Extra targets limit the lengths of the longest
slits, which made for easier data reduction because the focal plane
curvature is not noticeable for short slits.

In principle, any selection in the CMD could impose a bias in the age
and metallicity distributions derived for the stars.  For example, if
the bounding isochrone on the red is too stringent, the selection will
exclude metal-rich and/or old stars.  In practice, our selection box
is generous.  The color selection likely does not impose any
significant bias in the age and metallicity distributions except to
exclude very young stars that have not reached or will never reach the
RGB\@.  However, our sample may have a slight metallicity bias due to
the magnitude limit.  The magnitudes of stars near the top of the RGB
are especially sensitive to metallicity.  Metal-poor stars are
brighter.  Therefore, some of the most-metal rich stars in our sample
will be fainter and have spectra with lower S/N than metal-poor stars
at the same evolutionary phase, leading to slight bias against
metal-rich stars.  The effect would not bias our results beyond the
error bars that we quote.

Figures~\ref{fig:leoa}--\ref{fig:sagdig} show the CMDs and coordinate
maps of the galaxies.  Stars that we identified to be members of their
respective galaxies (Section~\ref{sec:membership}) are shown as blue
points.  Stars that failed any of the membership criteria are shown as
red crosses.  This category also includes stars whose spectra were so
noisy that we could not measure a radial velocity.  Small black points
identify stars for which we did not obtain a spectrum.

\subsection{Observations}

\begin{deluxetable*}{llccccccc}
\tablewidth{0pt}
\tablecolumns{9}
\tablecaption{DEIMOS Observations\label{tab:obs}}
\tablehead{\colhead{Galaxy} & \colhead{Slitmask} & \colhead{Targets} & \colhead{Slit width} & \colhead{Tot.\ Exp.\ Time} & \colhead{Exp.\ Time} & \colhead{Exposures} & \colhead{Seeing} & \colhead{UT Date} \\
\colhead{ } & \colhead{ } & \colhead{ } & \colhead{($''$)} & \colhead{(hr)} & \colhead{(min)} & \colhead{ } & \colhead{($''$)} & \colhead{ }}
\startdata
Leo A    & leoaaW\tablenotemark{a} &    121 & 1.1 & \phn6.7 &    400 &    14 &     0.9 & 2013 Jan 14    \\
         & leoA                    & \phn75 & 0.7 & \phn4.0 &    240 &    12 &     1.0 & 2013 Apr 1\phn \\
         & leoac                   &    120 & 0.7 & \phn5.5 &    330 &    11 &     1.0 & 2014 Feb 2\phn \vspace{0.25cm} \\
Aquarius & aqra\tablenotemark{a}   & \phn77 & 0.7 & \phn8.9 &    333 &    12 &     0.5 & 2013 Jul 8\phn \\
         &                         &        &     &         & \phn60 & \phn2 &     0.7 & 2013 Sep 1\phn \\
         &                         &        &     &         &    142 & \phn5 &     0.9 & 2013 Sep 2\phn \\
         & aqrd                    & \phn73 & 1.1 & \phn4.3 &    259 & \phn9 &     1.4 & 2013 Jul 9\phn \vspace{0.25cm} \\
SagDIG   & sagdia                  & \phn88 & 0.7 & \phn8.9 &    294 &    10 &     0.8 & 2014 Jun 29    \\
         &                         &        &     &         &    240 & \phn8 &     0.7 & 2014 Aug 28    \\
         & sagdib                  & \phn85 & 0.7 &    11.0 &    330 &    11 &     0.8 & 2014 Jun 30    \\
         &                         &        &     &         & \phn30 & \phn1 &     0.9 & 2014 Aug 29    \\
         &                         &        &     &         &    150 & \phn5 &     0.9 & 2014 Aug 30    \\
         &                         &        &     &         &    150 & \phn5 &     0.6 & 2014 Aug 31    \\
\enddata
\tablenotetext{a}{Observations published by \citet{kir13b}.}
\end{deluxetable*}

We observed the three galaxies with DEIMOS over several nights in 2013
and 2014.  We used the 1200G grating, which has a groove spacing of
1200~mm$^{-1}$ and a blaze wavelength of 7760~\AA\@.  We set the
central wavelength to 7800~\AA, and we used the OG550 order-blocking
filter.  This configuration provides a resolving power of $R \sim
7000$ at the central wavelength.  We obtained images of an internal
quartz lamp for flat fielding and internal Ne, Ar, Kr, and Xe arc
lamps for wavelength calibration.  Table~\ref{tab:obs} lists the dates
and conditions of the observations.  This paper includes two
slitmasks, leoaaW and aqra, that were previously published by
\citet{kir13b}.

We reduced the data with the spec2d pipeline \citep{coo12,new13}.  The
pipeline extracts rectangular areas of the CCD image corresponding to
the spectrally dispersed images of the slit.  Each slit image is flat
fielded and wavelength calibrated.  The wavelength calibration is
based on the arc lamp images and refined by the measured pixel
positions of terrestrial sky emission lines.  The stellar spectrum is
automatically identified and extracted with optimal weighting.  We
incorporated some modifications to spec2d that improve the wavelength
calibration and account for the curvature of the 2-D spectrum due to
differential atmospheric refraction \citep{kir15a,kir15b}.  The
pipeline tracks the variance of the spectrum so that the final 1-D
spectrum has both flux and an error on the flux.

The slitmasks sagdia, sagdib, and aqra had observations separated by
more than a month.  For these slitmasks, we grouped together the
observations taken in the same week and reduced them together.  That
left us with two independent sets of reduced 1-D spectra from the same
slitmask.  We applied a heliocentric correction to the wavelength
array of the second reduction to bring it into the same velocity
reference frame as the first reduction.  Then, we coadded the two sets
of spectra, weighting each pixel by its inverse variance.
Furthermore, some stars were observed on multiple slitmasks.  We
coadded their spectra in the same manner in order to achieve the best
S/N for each star.

\begin{figure}[t!]
\centering
\includegraphics[width=\columnwidth]{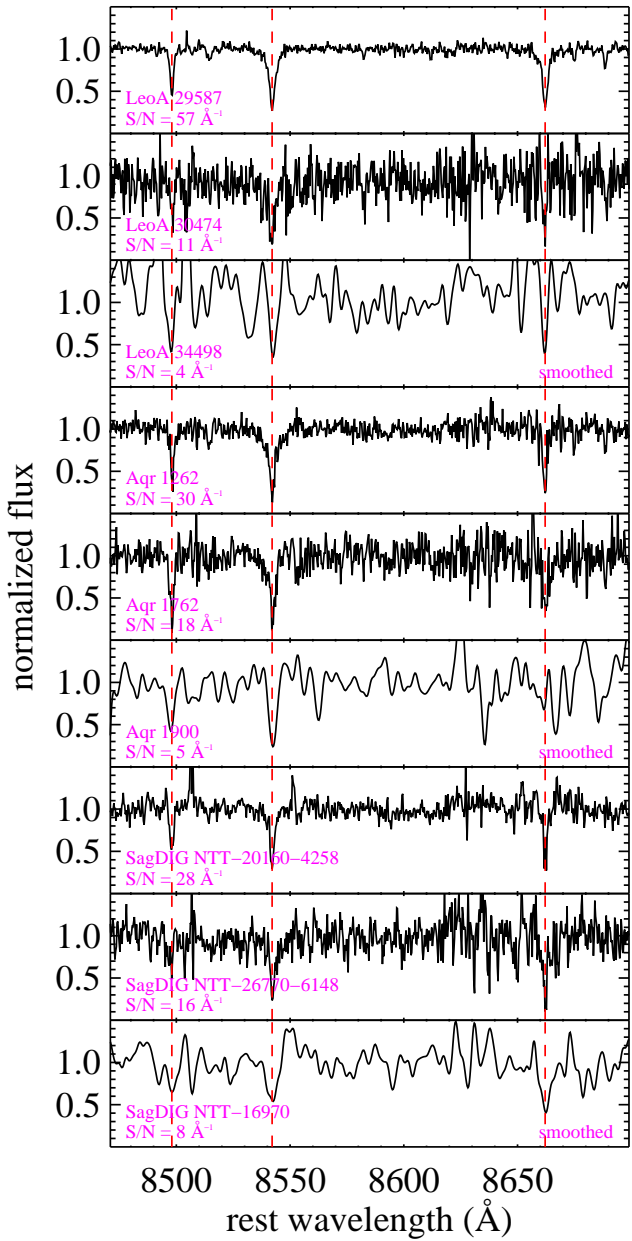}
\caption{DEIMOS spectra with the highest, median, and lowest S/N of
  the member stars for each of Leo~A, Aquarius, and SagDIG\@.  Red
  dashed lines indicate the rest wavelengths of the Ca$\,${\sc ii}
  triplet.  For display only, the spectra with the lowest S/N are
  smoothed to reduce noise.  Each panel gives the name of each star
  and the S/N of the spectrum.\label{fig:spectra}}
\end{figure}

Figure~\ref{fig:spectra} shows example spectra in each of the
galaxies.  The spectra were chosen to illustrate the full range of S/N
in our data set.  The topmost spectrum, LeoA~29587, shows an example
that was coadded from two different slitmasks (leoaaW and leoac).


\section{Spectroscopic Measurements}
\label{sec:meas}

We measured radial velocities and metallicities from the DEIMOS
spectra of individual stars in the three dIrrs.  We used the radial
velocities to help determine whether each targeted star was a member
of the galaxy.

\subsection{Radial Velocities}
\label{sec:vel}

Our technique to measure the radial velocities \citep[based
  on][]{sim07} is the same as we have used before.  The technique is
based on matching the observed spectrum to empirical templates.  This
process is similar to the DEEP2 collaboration's galaxy redshift
measurement technique \citep{new13}.  Specifically, the velocity
$v_{\rm obs}$ is measured by minimizing $\chi^2$ between the observed
spectrum and a template spectrum.  The template with the lowest
$\chi^2$ is the one used for the final velocity measurement.  We used
the nine metal-poor stars observed as radial velocity standards by
\citet{kir15a}.  We transformed all velocities into the heliocentric
frame.

The observed velocity of the star can vary with the position of the
star perpendicular to the slit.  Ideally, the star will be centered in
the slit, but misalignment can translate into a spurious velocity
offset of a few km~s$^{-1}$ with respect to the arc lines or sky
emission lines, which fill the slit.  The offset can be corrected by
establishing a geocentric frame of reference based on the telluric
absorption lines \citep{soh07,sim07}.  However, some of our spectra
span observations separated by months.  The heliocentric velocity
zeropoint shifts significantly over that time.  As a result, the
telluric lines in the stacked spectra are smeared over a range of
wavelengths, precluding a clean measurement of the slit mis-centering
correction.  Consequently, we did not perform this correction on any
of our data.  Regardless, the correction is much less important for
dIrrs than ultra-faint dSphs \citep[e.g.,][]{sim07,kir15c,kir15a}
because the the typical magnitude of the correction is
2.5~km~s$^{-1}$, much smaller than the velocity dispersions of dIrrs.
Furthermore, the correction becomes more important for observations
where the seeing drops below the slit width.  That was not the case
for most of our observations.

We estimated random velocity uncertainty by resampling the observed
spectrum 1000 times.  In each Monte Carlo trial, we added noise in the
spectrum.  The added noise in each pixel was sampled from a Gaussian
random distribution with a width equal to the square root of the
variance on that pixel's flux.  Hence, each pixel was perturbed on
average by $1\sigma$ from its nominal value.  We re-measured the
velocity from the noise-added spectrum, considering only the radial
velocity template that best matched the original spectrum.  The random
uncertainty, $\delta_{\rm rand} v$, is the standard deviation of
$v_{\rm helio}$ measured from all 1000 trials.  \citet{sim07} found
that this estimate of velocity uncertainty was too small for spectra
with high signal-to-noise ratios (S/N).  They calculated a systematic
error, $\delta_{\rm sys} v$, based on repeat measurements of the same
stars.  \citet{kir15a} also followed that procedure to determine
$\delta_{\rm sys} v = 1.49$~km~s$^{-1}$.  The final error on each
measurement is $\delta v = \sqrt{\delta_{\rm sys} v^2 + \delta_{\rm
    sys} v^2}$.

\subsection{Chemical Abundances}
\label{sec:chemical}

We measured metallicities and some detailed abundance ratios by
spectral synthesis.  We used the same technique as
\citet{kir08a,kir10}, who give more details than provided here.
Specifically, we compared the observed spectra with a large grid of
synthetic spectra spanning the expected ranges for old red giants of
effective temperature, surface gravity, metallicity, and alpha element
enhancement.

We prepared the spectra for abundance measurements first by dividing
by the spectrum of a hot star to correct for telluric absorption.
Then, we fit a polynomial to line-free regions of the spectrum.  This
polynomial was the first attempt at determining the stellar continuum.
We then used Levenberg--Marquardt minimization to search the grid of
synthetic spectra for the model spectrum with the minimum $\chi^2$
when compared with the observed, continuum-normalized spectrum.  We
fixed the surface gravity at the value determined by fitting model
isochrones to the star's broadband magnitude and color.  The
temperature, [Fe/H], and [$\alpha$/Fe] were free parameters, although
the temperature was constrained by both the spectrum and by the
broadband photometry.  After determining the best-fit atmospheric
parameters, we fit a new continuum polynomial to the quotient of the
observed spectrum and the best-fitting model.  We continually re-fit
the spectrum with successive refinements of the continuum until the
procedure converged.  Finally, we measured individual abundances of
Mg, Si, Ca, and Ti\footnote{We adopted solar abundances of
  $\epsilon({\rm Fe}) \equiv 12 + \log [n({\rm Fe})/n({\rm H})] =
  7.52$ \citep{sne92}, $\epsilon({\rm Mg}) = 7.58$, $\epsilon({\rm
    Si}) = 7.55$, $\epsilon({\rm Ca}) = 6.36$, $\epsilon({\rm Ti}) =
  4.99$ \citep{and89}.} by restricting the spectral fit to regions of
the spectrum that contain absorption lines of each of those elements.

We estimated random uncertainties on the abundances from the diagonal
elements of the covariance matrix produced by the Levenberg--Marquardt
grid search.  Like the radial velocity uncertainties, the random
uncertainty is an incomplete estimate of the error.  We computed the
total error on an element's abundance by adding a systematic error in
quadrature with the random error.  \citet{kir10} determined the
magnitude of the systematic errors by comparing repeated measurements
of the same stars.  The systematic errors are 0.106~dex for [Fe/H],
0.065~dex for [Mg/Fe], 0.113~dex for [Si/Fe], 0.111~dex for [Ca/Fe],
and 0.090~dex for [Ti/Fe].

\subsection{Membership}
\label{sec:membership}

Not all stars that we observed belong to the dIrr galaxies that we are
studying.  We required every star to pass three membership criteria:
(1) position in the CMD, (2) absence of spectral features that would
indicate non-membership, and (3) radial velocity.

First, the stars need to have the approximate colors and magnitudes of
red giants at the distance of each galaxy.  Although we prioritized
the design of the DEIMOS slitmasks to target such stars, we also
targeted some stars that could not be red giant members based on their
position in the CMD\@.  Any star that did not fall within the bounds
of the extreme red and blue isochrones described in
Section~\ref{sec:target} was excluded.

Second, we examined each spectrum for spectral features that indicate
that the star could not be a red giant member of the galaxy.  The most
obvious such features were redshifted emission lines, such as
H$\alpha$ or [O$\,${\sc ii}]$\,\lambda 3727$.  These features
indicated that the target was a background galaxy.  Another telltale
spectral feature was the Na$\,${\sc i}$\,\lambda8190$ doublet.  This
feature is very strong in dwarf stars.  At the distances of dIrrs,
dwarf stars would be much fainter than the apparent magnitudes of our
target stars.  Hence, a strong Na doublet indicates that the star is
in the foreground \citep{spi71,coh78}.  We measured equivalent widths
by fitting Gaussians for weaker doublets and Lorentzians for stronger
doublets.  \citet{kir12a} found that a combined equivalent width of
the two Na lines stronger than 1~\AA\ indicates that the star is a
dwarf for any reasonable range of Na abundance.  We adopted the same
membership criterion.

We also noted the presence of TiO and CN bands.  TiO is generally
found in cool, metal-rich stars.  Hence, the presence of TiO is a
likely indicator of non-membership, but it is not definitive.  Hence,
we did not use it to determine membership.  Regardless, every star
with visible TiO bands failed at least one other membership criterion.
CN bands usually indicate that a star is enhanced in carbon.  Carbon
stars on the RGB or asymptotic giant branch (AGB) are not particularly
rare, even in metal-poor dwarf galaxies.  Hence, we did not use the
presence of CN as a membership indicator.  However, we did exclude any
stars with obvious CN from the measurement of chemical abundances
because CN lines are not adequately represented in our spectral
syntheses.

\begin{figure*}[t!]
\centering
\includegraphics[width=0.95\textwidth]{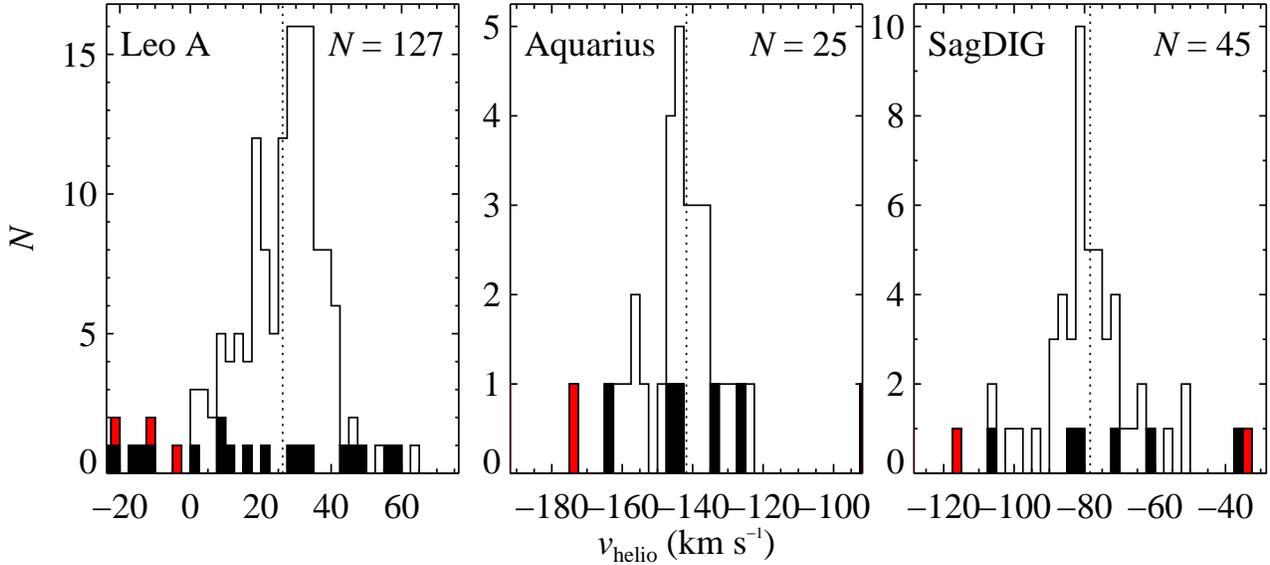}
\caption{Histograms of radial velocities in 2.5~km~s$^{-1}$ bins.
  Color coding indicates non-membership
  (Section~\ref{sec:membership}).  Red shading indicates stars ruled
  as non-members by their velocities.  Black shading indicates stars
  that are ruled out by their position in the CMD or by the presence
  of a strong Na~{\sc i}~8190 doublet in their spectra.  The vertical
  dotted line indicates the mean radial velocity.  The upper right
  corners show the number of member stars.\label{fig:vhist}}
\end{figure*}

Third, we identified member stars on the basis of their radial
velocities.  We followed the same process as our previous work on the
kinematics of dwarf galaxies \citep{kir14,kir15a}.  We required that
member stars have radial velocities that fall within $2.58\sigma_v$ of
the mean velocity, $\langle v_{\rm helio} \rangle$, where $\sigma_v$
is the velocity dispersion (Section~\ref{sec:kinematics}).  This
criterion includes 99\% of the member stars if their velocities are
normally distributed.  However, some of the stars have velocity errors
on the order of $\sigma_v$.  To accommodate these stars, we also count
as members all stars whose $1\sigma$ velocity uncertainty overlaps the
$\pm 2.58\sigma_v$ velocity range.  Put concisely, the velocity
requirement for membership is $|v_{\rm helio} - \langle v_{\rm helio}
\rangle| - \delta v < 2.58\sigma_v$.

Figure~\ref{fig:vhist} shows the velocity distributions of stars near
the mean velocity for each dIrr.  The range of the plots include all
member stars but not all non-members.  Black shading indicates
non-members that were excluded because of their CMD position or the
presence of a strong Na doublet.  Red shading indicates stars that
were excluded only on the basis of radial velocity.  Background
galaxies are not shown.

\begin{deluxetable*}{llcccccrcll}
\tablewidth{0pt}
\tablecolumns{11}
\tablecaption{Radial Velocities and Membership\label{tab:velocity}}
\tablehead{\colhead{Galaxy} & \colhead{ID} & \colhead{RA (J2000)} & \colhead{Dec (J2000)} & \colhead{$I_0$} & \colhead{$(V-I)_0$} & \colhead{Masks} & \colhead{S/N\tablenotemark{a}} & \colhead{$v_{\rm helio}$} & \colhead{Member?} & \colhead{Flags\tablenotemark{b}} \\
\colhead{ } & \colhead{ } & \colhead{ } & \colhead{ } & \colhead{(mag)} & \colhead{(mag)} & \colhead{ } & \colhead{(\AA$^{-1}$)} & \colhead{(km~s$^{-1}$)} & \colhead{ } & \colhead{ }}
\startdata
LeoA     & 29115            & 09 59 03.87 & $+30$ 46 32.6 & 18.64 & $ 2.29$ & 1 &     \phn  97 & $   0.1 \pm  1.5$ & N & TiO v CMD Na \\
LeoA     & 31326            & 09 59 04.81 & $+30$ 48 08.9 & 21.89 & $ 1.10$ & 3 &     \phn  15 & $  32.2 \pm  3.1$ & Y &              \\
LeoA     & 9282             & 09 59 05.57 & $+30$ 45 25.9 & 21.51 & $ 1.11$ & 1 & \phn\phn   6 & $  35.0 \pm  5.6$ & Y & CN           \\
LeoA     & 11200            & 09 59 05.64 & $+30$ 46 16.2 & 21.60 & $ 1.20$ & 1 & \phn\phn   7 & $  37.7 \pm  5.1$ & Y &              \\
LeoA     & 11058            & 09 59 06.22 & $+30$ 46 10.9 & 21.70 & $ 1.18$ & 1 & \phn\phn   6 & $  38.5 \pm  7.9$ & Y &              \\
LeoA     & 30011            & 09 59 06.33 & $+30$ 45 56.0 & 21.02 & $ 1.29$ & 2 &     \phn  31 & $  16.7 \pm  2.0$ & Y &              \\
LeoA     & 34130            & 09 59 06.87 & $+30$ 46 49.5 & 22.90 & $ 1.15$ & 2 & \phn\phn   6 & $  37.1 \pm 34.7$ & Y &              \\
LeoA     & 31582            & 09 59 07.09 & $+30$ 45 36.5 & 22.27 & $ 0.98$ & 2 & \phn\phn   7 & $  28.3 \pm  6.1$ & Y &              \\
LeoA     & 31580            & 09 59 07.78 & $+30$ 45 17.2 & 22.05 & $ 1.02$ & 2 &     \phn  12 & $  11.5 \pm  3.5$ & Y &              \\
LeoA     & 9397             & 09 59 07.90 & $+30$ 45 28.2 & 21.11 & $ 1.30$ & 1 &     \phn  12 & $  17.8 \pm  3.2$ & Y &              \\
\nodata & \nodata & \nodata & \nodata & \nodata & \nodata & \nodata & \nodata & \nodata & \nodata & \nodata \\
\enddata
\tablenotetext{a}{To convert to S/N per pixel, multiply by 0.57.}
\tablenotetext{b}{galaxy: Spectrum indicates that the object is a galaxy, not an individual star.  TiO: TiO bands present in spectrum (not necessarily an indicator of membership).  CN: Spectrum shows strong CN features (not an indicator of non-membership).  v: Non-member by radial velocity.  CMD: Non-member by location in CMD\@.  Na: Non-member by presence of strong Na$\,${\sc i}$\,$8190 doublet.}
\tablecomments{(This table is available in its entirety in machine-readable form.)}
\end{deluxetable*}

\begin{deluxetable*}{llcccccccc}
\tablewidth{0pt}
\tablecolumns{10}
\tablecaption{Chemical Abundances\label{tab:metallicity}}
\tablehead{\colhead{Dwarf} & \colhead{ID} & \colhead{$T_{\rm eff}$} & \colhead{$\log g$} & \colhead{[Fe/H]} & \colhead{[Mg/Fe]} & \colhead{[Si/Fe]} & \colhead{[Ca/Fe]} & \colhead{[Ti/Fe]} & \colhead{[$\alpha$/Fe]} \\
\colhead{ } & \colhead{ } & \colhead{(K)} & \colhead{(cm~s$^{-2}$)} & \colhead{ } & \colhead{ } & \colhead{ } & \colhead{ } & \colhead{ } & \colhead{ }}
\startdata
LeoA     & 33186            & 4584 & 1.54 & $-1.82 \pm 0.30$ &     \nodata      &     \nodata      &     \nodata      &     \nodata      &     \nodata      \\
LeoA     & 29857            & 4299 & 0.63 & $-2.51 \pm 0.12$ &     \nodata      & $+0.85 \pm 0.21$ & $+0.21 \pm 0.29$ & $+0.73 \pm 0.25$ & $+0.09 \pm 0.33$ \\
LeoA     & 30093            & 4303 & 0.80 & $-0.57 \pm 0.11$ &     \nodata      &     \nodata      &     \nodata      & $-0.47 \pm 0.31$ & $-0.77 \pm 0.28$ \\
LeoA     & 30073            & 4295 & 0.76 & $-2.02 \pm 0.11$ &     \nodata      & $+0.47 \pm 0.21$ & $-0.03 \pm 0.25$ & $+0.35 \pm 0.15$ & $+0.11 \pm 0.17$ \\
LeoA     & 31326            & 4617 & 1.23 & $-1.74 \pm 0.13$ & $+1.12 \pm 0.44$ &     \nodata      & $+0.07 \pm 0.39$ &     \nodata      &     \nodata      \\
LeoA     & 11200            & 4432 & 1.04 & $-1.41 \pm 0.19$ &     \nodata      &     \nodata      &     \nodata      &     \nodata      &     \nodata      \\
LeoA     & 11058            & 4461 & 1.10 & $-1.49 \pm 0.22$ &     \nodata      &     \nodata      &     \nodata      &     \nodata      &     \nodata      \\
LeoA     & 30011            & 4340 & 0.77 & $-1.93 \pm 0.12$ & $+0.84 \pm 0.42$ &     \nodata      & $+0.08 \pm 0.36$ & $+0.07 \pm 0.20$ & $-0.00 \pm 0.23$ \\
LeoA     & 31582            & 4879 & 1.47 & $-1.62 \pm 0.36$ &     \nodata      &     \nodata      &     \nodata      &     \nodata      &     \nodata      \\
LeoA     & 31580            & 4814 & 1.36 & $-1.30 \pm 0.15$ &     \nodata      &     \nodata      &     \nodata      &     \nodata      &     \nodata      \\
\nodata & \nodata & \nodata & \nodata & \nodata & \nodata & \nodata & \nodata & \nodata & \nodata \\
\enddata
\tablecomments{(This table is available in its entirety in machine-readable form.)}
\end{deluxetable*}

Table~\ref{tab:velocity} gives the names, coordinates,
extinction-corrected $I_0$ magnitude, reddening-corrected $(V-I)_0$
color, number of slitmasks on which the star was observed, S/N, radial
velocity, and membership (yes or no).  The last column indicates
various reasons for non-membership or other qualities of the spectrum,
such as the presence of CN absorption in member stars.  The table
contains \nmem\ member stars and \nnonmem\ non-members for all three
dIrrs.  Table~\ref{tab:metallicity} gives temperatures, gravities, and
abundances for the \nfeh\ stars where those values were measurable.
The last column, [$\alpha$/Fe], is not an arithmetic average of the
previous four columns.  Rather, it is a determination of the
[$\alpha$/Fe] ratio using all of the available $\alpha$ element lines
in the spectrum.  In this sense, it is an average weighted by the S/N
of the relevant Mg, Si, Ca, and Ti absorption lines.  This is the
value called $[\alpha/{\rm Fe}]_{\rm atm}$ by \citet{kir10}.

We compared our membership list for SagDIG with \citeauthor{mom14}'s
(\citeyear{mom14}) proper motion-culled catalog.  The field of view of
HST, from which the proper motions were measured, limits the overlap
between that catalog and our DEIMOS catalog to \sagdigoverlap\ stars.
Of them, we classified \sagdigmommem\ as members, all of which have
proper motions less than $1.2$~mas~yr$^{-1}$.  This value is within
the approximate range that \citeauthor{mom14}\ considered stars to
pass the proper motion cut.  Of the eight stars that we classified as
non-members, only one had a proper motion in excess of
$1.2$~mas~yr$^{-1}$.  Thus, the proper motion cut does not misclassify
members as non-members, but it is not especially efficient at ruling
out non-members.


\section{Kinematics}
\label{sec:kinematics}

\subsection{Velocity Dispersion}

We measured the velocity dispersions of each dIrr using maximum
likelihood.  Our approach is similar to that of \citet{wal06} and
identical to that of \citet{kir15a}.  The likelihood, $L$, that the
galaxy has a mean velocity $\langle v_{\rm helio} \rangle$ and a
velocity dispersion $\sigma_v$ is

\begin{eqnarray}
\nonumber \ln L &=& -\frac{1}{2} \sum_i^N \ln \left[ 2\pi \left((\delta v)_i^2 + \sigma_v^2\right) \right] \\
& & - \frac{1}{2} \sum_i^N \left(\frac{\left((v_{\rm helio})_i - \langle v_{\rm helio} \rangle \right)^2}{\left(\delta v \right)_i^2 + \sigma_v^2}\right) \label{eq:v}
\end{eqnarray}

\noindent
The sum is performed over the velocity measurements of the $N$ unique
member stars in each dIrr.  The velocity and error on the $i^{\rm th}$
star are $(v_{\rm helio})_i$ and $(\delta v)_i$ (see
Section~\ref{sec:vel}).  We maximized the likelihood using a Monte
Carlo Markov chain (MCMC) with a Metropolis--Hastings algorithm.

The determination of $\langle v_{\rm helio} \rangle$ and $\sigma_v$
depends on the stars included in the member list.  However, the
velocity criterion for membership (Section~\ref{sec:membership})
depends on $\langle v_{\rm helio} \rangle$ and $\sigma_v$.  Therefore,
these two parameters must be determined iteratively.  First, we made
rough guesses at the mean velocity and velocity dispersion by
calculating the mean and standard deviation of possible members in the
velocity ranges shown in Figure~\ref{fig:vhist}.  These values were
used to construct the initial member list.  Second, we re-evaluated
$\langle v_{\rm helio} \rangle$ and $\sigma_v$ using
Equation~\ref{eq:v} and an MCMC chain with $10^5$ iterations.  These
values were used to refine the member list.  Third, we repeated this
process until the membership list remained the same from one iteration
to the next.  Finally, we re-evaluated $\langle v_{\rm helio} \rangle$
and $\sigma_v$ with the final member list using an MCMC chain with
$10^7$ iterations.  The difference between $10^5$ and $10^7$
iterations is not significant enough to alter the member list, but it
does help us better determine the confidence intervals on the
measurements.

Using a Markov chain allows us to sample the parameter space well
enough to determine one-sided confidence intervals.  We quote error
bars as the values that enclose 68.3\% of the MCMC trials.  The error
bars are allowed to be asymmetric, such that the parameter space
between the lower error bar and the mean includes 34.2\% of the
trials, and the parameter space between the mean and the upper error
bar includes another 34.2\% of the trials.

The dynamical mass of a galaxy can be estimated from its line-of-sight
velocity dispersion and projected half-light radius.  Although there
is some uncertainty from the unknown velocity anisotropy and from the
de-projection of the 2-D half-light radius to its 3-D value,
\citet{wol10} found a robust mass estimator that minimizes the effect
of these uncertainties.  The mass within the 3-D half-light radius is
$M_{1/2} = 4G^{-1}\sigma_v^2 r_h$, where $\sigma_v$ is the
line-of-sight velocity dispersion and $r_h$ is the 2-D half-light
radius.  These two quantities are directly measured from observations.

\begin{deluxetable*}{lcccc}
\tablewidth{0pt}
\tablecolumns{5}
\tablecaption{Galaxy Properties\label{tab:properties}}
\tablehead{\colhead{Property} & \colhead{Leo A} & \colhead{Aquarius} & \colhead{SagDIG} & \colhead{Unit}}
\startdata
\cutinhead{Photometric Properties}
Distance & $ 827 \pm  11$ (1) & $ 977 \pm 45$ (2) & $1047 \pm 53$ (3) & kpc \\
$L_V$ & $6.6 \pm 1.4$ (4) & $1.7 \pm 0.2$ (5) & $4.6 \pm 1.1$ (6) & $10^6~L_{\sun}$ \\
$r_h$ & $2.15 \pm 0.12$ (4) & $1.10 \pm 0.03$ (5) & $0.91 \pm 0.05$ (6) & arcmin \\
$r_h$ & $517 \pm  29$ (4) & $312 \pm  16$ (5) & $277 \pm  20$ (6) & pc \\
$M_*$ & $3.3 \pm 0.7$ (7) & $1.5 \pm 0.2$ (7) & $1.8 \pm 0.5$ (7) & $10^6~M_{\sun}$ \\
SFR(H$\alpha$) & 9.3 (8) & 0 (9,10) & 8.5 (8) & $10^{-5}~M_{\sun}$~yr$^{-1}$ \\
SFR(UV) & 6.0 (8) & 0 (11) & 7.2 (8) & $10^{-4}~M_{\sun}$~yr$^{-1}$ \\
\cutinhead{Gas Properties}
$M$(H$\,${\sc i}) & $7.4 \pm 0.8$ (12) & $2.2 \pm 0.3$ (12) & $8.3 \pm 1.2$ (12) & $10^6~M_{\sun}$ \\
$\langle v_{\rm helio} \rangle$ (H$\,${\sc i}) & $  23.7$ (12) & $-140.3$ (12) & $ -79.2$ (12) & km~s$^{-1}$ \\
$\sigma_v$ (H$\,${\sc i}) & $6.2$ (12) & $6.7$ (12) & $8.2$ (12) & km~s$^{-1}$ \\
\cutinhead{Stellar Dynamical Properties}
$N_{\rm member}$ & 127 &  25 &  45 & \\
$\langle v_{\rm helio} \rangle$ & $  26.2_{-0.9}^{+1.0}$ & $-141.8_{-2.0}^{+1.8}$ & $ -78.4 \pm 1.6$ & km~s$^{-1}$ \\
$v_{\rm GSR}$ & $ -13.9$ & $ -30.7$ & $   6.2$ & km~s$^{-1}$ \\
$\sigma_v$ & $ 9.0_{-0.6}^{+0.8}$ & $ 7.8_{-1.1}^{+1.8}$ & $ 9.4_{-1.1}^{+1.5}$ & km~s$^{-1}$ \\
$M_{1/2}$\tablenotemark{a} & $3.9 \pm 0.4$ & $1.8_{-0.3}^{+0.4}$ & $2.3_{-0.3}^{+0.4}$ & $10^7~M_{\sun}$ \\
$(M/L_V)_{1/2}$\tablenotemark{b} & $  12 \pm 3$ & $  21_{-   4}^{+   6}$ & $  10 \pm 3$ & $M_{\sun}~L_{\sun}^{-1}$ \\
$(M_{\rm tot}/M_b)_{1/2}$\tablenotemark{c} & $7.3_{-1.0}^{+1.1}$ & $9.6_{-1.8}^{+2.5}$ & $4.4_{-0.8}^{+1.0}$ & \\
\cutinhead{Stellar Chemical Properties}
$\langle {\rm [Fe/H]} \rangle$ & $-1.67_{-0.08}^{+0.09}$ & $-1.50 \pm 0.06$ & $-1.88_{-0.09}^{+0.13}$ & \\
$p_{\rm eff}$ (Leaky Box) & $3.5 \pm 0.4$ & $5.5_{-1.1}^{+1.2}$ & $2.4_{-0.4}^{+0.5}$ & $10^{-2}~Z_{\sun}$ \\
$p_{\rm eff}$ (Pre-Enriched) & $2.6_{-0.3}^{+0.4}$ & $5.3_{-1.2}^{+1.4}$ & $2.4_{-0.4}^{+0.5}$ & $10^{-2}~Z_{\sun}$ \\
${\rm [Fe/H]}_0$ (Pre-Enriched) & $ -2.47_{- 0.12}^{+ 0.10}$ & $< -2.28$\tablenotemark{d} & $< -3.99$\tablenotemark{d} & \\
$\Delta (\ln P)$ (Pre-Enriched)\tablenotemark{e} & $12.13$ & $-1.68$ & $-1.72$ & \\
$p_{\rm eff}$ (Accretion) & $3.2 \pm 0.3$ & $4.4_{-0.7}^{+0.8}$ & $2.1_{-0.3}^{+0.4}$ & $10^{-2}~Z_{\sun}$ \\
$M$ (Accretion) & $6.0_{- 2.0}^{+ 3.2}$ & $7.1_{- 3.8}^{+ 6.2}$ & $2.3_{- 0.9}^{+1.9}$ & \\
$\Delta (\ln P)$ (Accretion)\tablenotemark{e} & $11.03$ & $ 0.37$ & $-1.19$ & \\
\enddata
\tablerefs{(1) \citet{tam11}.  (2) \citet{col14}.  (3) \citet{mom05}.  $L_V$ and $r_h$ based on surface brightness profiles from (4) \citet{dev91}, (5) \citet{mcc06}, and (6) \citet{lee00}.  Both values are updated for the distances adopted here.  $L_V$ is corrected for extinction based on \citet{sch11}.  (7) Based on stellar mass-to-light ratios from \citet{woo08}.  (8) \citet{kar13}.  (9) \citet{van00}.  (10) \citet{hun04}.  (11) \citet{hun10}.  (12) Measured from the H$\,${\sc i} maps of \citet{hun12}.  H$\,${\sc i} masses updated for the distances adopted here.  The uncertainty in $M$(H$\,${\sc i}) incorporates error on the 21~cm flux---assumed to be 11\%---and uncertainty in distance.}
\tablenotetext{a}{Mass within the half-light radius, calculated as $M_{1/2} = 4G^{-1}\sigma_v^2 r_h$ \citep{wol10}.}
\tablenotetext{b}{Mass-to-light ratio within the half-light radius.}
\tablenotetext{c}{Total (dynamical) mass divided by baryonic mass ($M_b = M_* + M($H$\,${\sc i}$)$) within the half-light radius.  The calculation assumes that half of the stellar mass and half of the gas mass is located within the half-light radius.}
\tablenotetext{d}{Upper limit with 95\% confidence.}
\tablenotetext{e}{Logarithm of the strength with which the Pre-Enriched or Accretion Model is favored over the Leaky Box model.  Negative values of $\Delta (\ln P)$ indicate that the model is disfavored.}
\end{deluxetable*}

Table~\ref{tab:properties} gives our measurements and confidence
intervals for $\langle v_{\rm helio} \rangle$, $\sigma_v$, and
$M_{1/2}$.  All of the galaxies have velocity dispersions in excess of
what would be expected from stellar mass alone.  The $V$-band
mass-to-light ratios of the known matter (stars and H$\,${\sc i} gas)
range from 1.6 to $2.2~M_{\sun}~L_{\sun}^{-1}$, but the observed
dynamical mass-to-light ratios range from \sagdigml\ to
$\aqrml~M_{\sun}~L_{\sun}^{-1}$.  Thus, dark matter outweighs luminous
matter by at least a factor of four.  This finding is consistent with
the velocity dispersions of H$\,${\sc i} gas
\citep[e.g.,][]{lo93,you96}.  However, the stars are not subject to
hydrodynamical pressure.  Thus, we can conclusively rule out gas
dynamics as the origin of the velocity dispersion.  Instead, the
galaxies must be dark matter-dominated.

\subsection{Rotation}
\label{sec:rot}

The tidal stirring model \citep{may01} posits that rotationally
supported dIrrs transform into dispersion-supported dSphs in the tidal
field of a massive host galaxy.  However, \citet{whe15} found that
stellar rotation is not common among dIrrs.  The lack of rotation
lessens the need for tides to transform a dIrr's dynamical support
from rotation to dispersion.  \citeauthor{whe15}\ already established
that Leo~A and Aquarius have limited rotation based on
\citeauthor{kir14}'s (\citeyear{kir14}) data.  However, the limit on
Aquarius was not particularly stringent due to the limited S/N of the
spectra.  In this section, we test for stellar rotation in our current
spectroscopic samples.

\begin{figure*}[t!]
\centering
\includegraphics[width=0.95\textwidth]{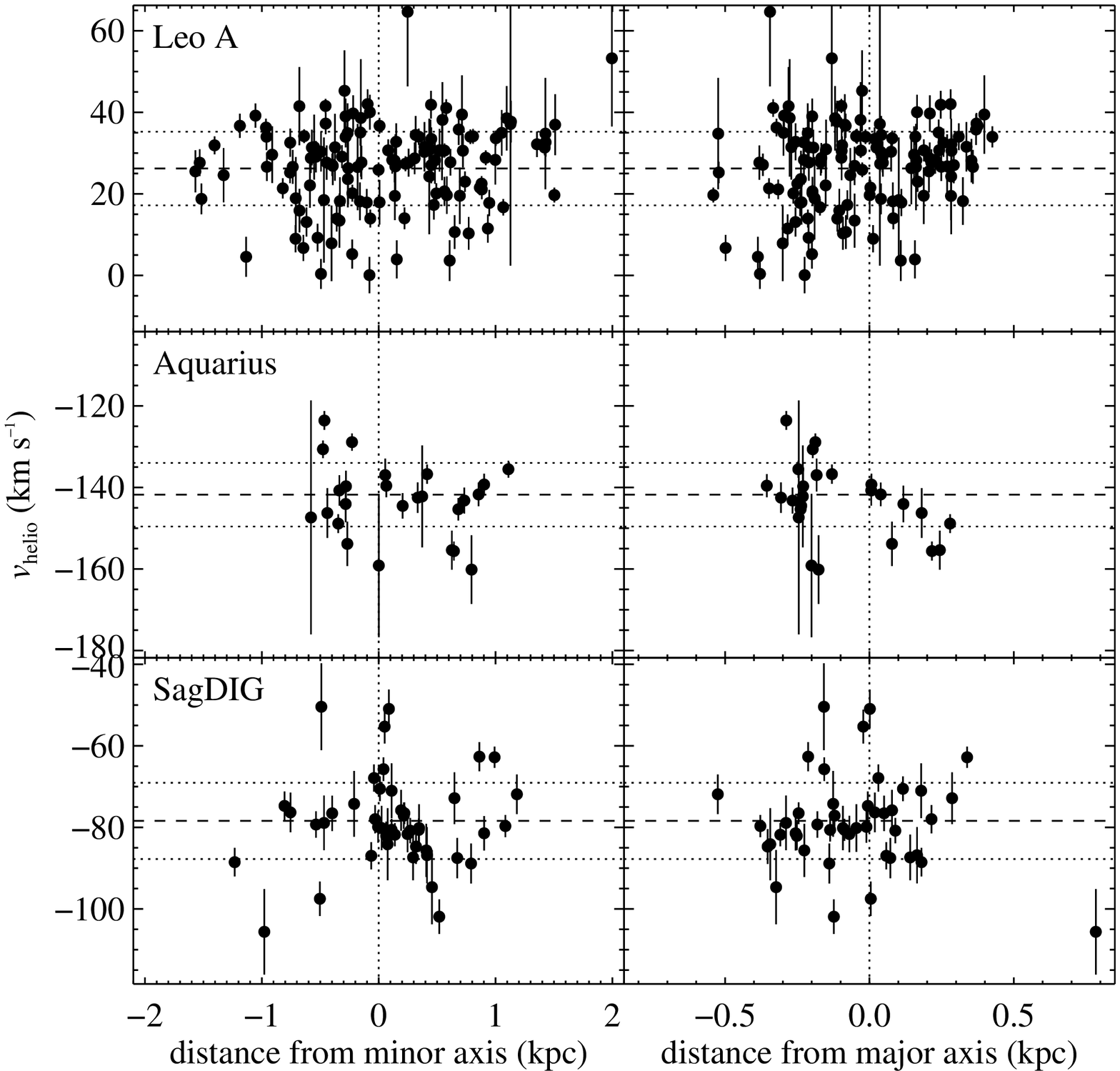}
\caption{The heliocentric radial velocities of member stars as a
  function of distance from the minor axis (left) and major axis
  (right) of their respective galaxy.  Rotation about the minor axis
  is more common.  However, stellar rotation is not apparent in either
  projection.  The horizontal dashed lines show the mean velocities,
  and the horizontal dotted lines show the $1\sigma$ velocity
  dispersions.\label{fig:rot}}
\end{figure*}

The most straightforward way to test for rotation is to plot the
stars' velocities versus their distances from the minor axis.  If the
galaxy is a rotationally supported disk, then the average velocity on
one side of the minor axis should be blueshifted relative to the mean
velocity, and the velocity on the other side should be redshifted.
The left panels of Figure~\ref{fig:rot} show this diagnostic.  No
rotation is apparent.  The average velocities on either side of the
minor axis seem about the same.  Given that some galaxies have been
found to exhibit prolate rotation \cite[e.g.,
  Andromeda~II,][]{ho12}---that is, rotation around the isophotal
major axis---Figure~\ref{fig:rot} also shows the velocities versus
their distances from the major axis.  This type of rotation is not
common, and indeed, we do not observe in any of the three dIrrs.

We also tested for rotation in a statistically rigorous manner.
Following \citet{whe15}, we modeled the velocity of each star with a
rotation term: $\langle v_{\rm helio} \rangle + v_{\rm rot} \cos
(\theta - \theta_i)$.  The magnitude of rotation is $v_{\rm rot}$, and
the axis of rotation is defined by the position angle $\theta$.  The
position angle of each star is $\theta_i$.  This parametrization
allows the rotation to occur around any axis, not just the minor axis.
We found the maximum likelihood values for $v_{\rm rot}$, $\theta$,
$\langle v_{\rm helio} \rangle$, and $\sigma_v$.  For computational
efficiency, we maximized the logarithm of the likelihood:

\begin{eqnarray}
\nonumber \ln L = -\frac{1}{2} \sum_i^N \ln \left[ 2\pi \left((\delta v)_i^2 + \sigma_v^2\right) \right] \\
- \frac{1}{2} \sum_i^N \left(\frac{\left[(v_{\rm helio})_i - \left(\langle v_{\rm helio} \rangle + v_{\rm rot} \cos (\theta - \theta_i) \right) \right]^2}{\left(\delta v \right)_i^2 + \sigma_v^2}\right) \label{eq:rot}
\end{eqnarray}

Equation \ref{eq:rot} is a generalization of Equation~\ref{eq:v}.  In
order to determine $v_{\rm rot}$ and $\theta$ freely and without bias,
we did not constrain $\langle v_{\rm helio} \rangle$ or $\sigma_v$ to
match the values we previously determined.  Regardless, the final
values computed from Equation~\ref{eq:rot} were indistinguishable from
those computed from Equation~\ref{eq:v}.  As before, we maximized the
likelihood with an MCMC chain with $10^5$ links.

We did not detect rotation in the dIrrs.  We constrained the rotation
to be $v_{\rm rot} < \leoavrotlim$, $\aqrvrotlim$, and
$\sagdigvrotlim$~km~s$^{-1}$ with 95\% confidence in Leo~A, Aquarius,
and SagDIG, respectively.  The corresponding limits on the ratio of
rotation velocity to velocity dispersion are $v_{\rm rot}/\sigma_v <
\leoavsigmalim$, $\aqrvsigmalim$, and $\sagdigvsigmalim$\@.  This
result is consistent with the results of \citet{whe15}.  Thus, we have
tightened the constraints on rotation in Leo~A and Aquarius and added
SagDIG to the list of non-rotating dIrrs.

\subsection{Comparison of the Stellar and Gas Kinematics}

Local Group dIrrs are the nearest gas-rich galaxies.  As such, they
have been observed fairly extensively with 21~cm radio measurements.
All three of Leo~A, Aquarius, and SagDIG contain copious amounts of
gas with gas-to-stellar mass ratios of $\leoags \pm \leoagserr$,
$\aqrgs \pm \aqrgserr$, and $\sagdiggs \pm \sagdiggserr$
\citep{hun12}.  It is interesting to compare the kinematics of the gas
with the kinematics of the stars.  The gas traces the recent history
of the galaxy, as well as hydrodynamical effects.  On the other hand,
the stars retain a longer dynamical memory than the gas, and they are
unaffected by collisional pressure.

\begin{figure*}[t!]
\centering
\includegraphics[width=\textwidth]{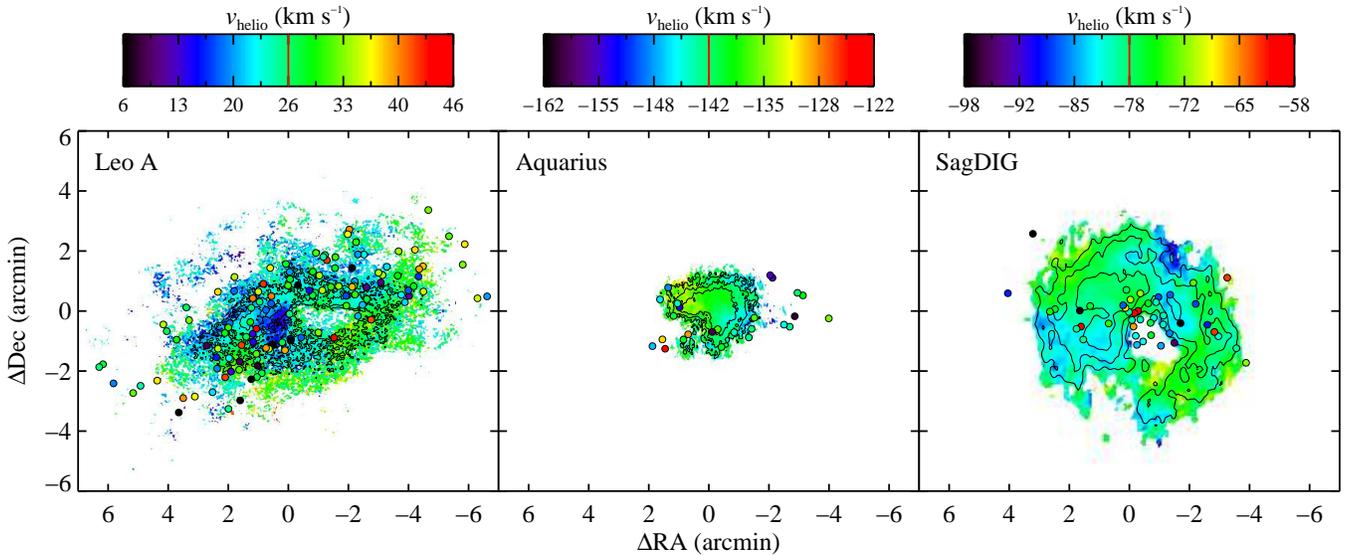}
\caption{The kinematics of H$\,${\sc i} gas in Leo~A, Aquarius, and
  SagDIG, as observed by LITTLE THINGS \protect \citep{hun12}.  The
  gas distribution is color-coded according to the velocity scale
  shown at the top of each plot.  A red line in the velocity scale
  indicates the average velocity, $\langle v_{\rm helio} \rangle$, of
  the stars.  The black contours indicate the flux of the 21~cm
  measurements.  The contour levels are 4, 8, and
  16~$M_{\sun}$~pc$^{-2}$ for Leo~A and 1.25, 2.5, and
  5~$M_{\sun}$~pc$^{-2}$ for Aquarius and SagDIG\@. The stars are
  shown as black-outlined circles.  The color of the circle shows the
  star's velocity on the same color scale as the gas.\label{fig:hi}}
\end{figure*}

We compared our measurements of stellar velocities with observations
of the H$\,${\sc i} gas by Local Irregulars that Trace Luminosity
Extremes: The H$\,${\sc i} Nearby Galaxy Survey \citep[LITTLE
  THINGS,][]{hun12}.  The 21~cm observations come from the Very Large
Array (VLA)\@.  Figure~\ref{fig:hi} shows the robust-weighted galaxy
maps with velocity color coding for both the stars and gas.  The
figure also shows the H$\,${\sc i} surface density contours.

For each galaxy, we computed the mean, flux-weighted velocity from the
robust-weighted VLA maps: $\langle v_{\rm helio} \rangle ($H$\,${\sc
  i}$) = \sum{f({\rm H})v({\rm H})}/\sum{f({\rm H})}$, where $f({\rm
  H})$ is the H$\,${\sc i} surface density in a pixel, and $v({\rm
  H})$ is the velocity from the intensity-weighted, first-moment map.
We computed a similar flux-weighted velocity dispersion: $\sigma_v
($H$\,${\sc i}$) = \sum{f({\rm H})\sigma_v({\rm H})}/\sum{f({\rm
    H})}$.  In this case, $\sigma_v({\rm H})$ comes from the
intensity-weighted, second-moment map.  Table~\ref{tab:properties}
shows those values, which can be compared to the stellar velocities in
the same table.  In general, the mean gas velocities agree with the
mean stellar velocities.  The gas velocity dispersions are slightly
less than the stellar velocity dispersions.  However, the gas
velocities explicitly exclude rotation.  Low levels of stellar
rotation or even stellar binaries could slightly inflate the stellar
velocity dispersions.

Leo~A, like all the dIrrs presented here, shows no sign of stellar
rotation or any velocity structure.  On the other hand, the H$\,${\sc
  i} velocity structure is more complex.  Although there is no obvious
rotation, there is some structure in the surface density and the
velocity field.  The majority of gas appears in an arc just north of
the galaxy's center.  The two sides of the arc appear as two lobes of
gas.  The velocity varies along the arc, but not in a way that
indicates rotation.  A blob of gas just east of Leo~A's center is
blueshifted relative the mean velocity.  Furthermore, there is an area
just southwest of the center that is devoid of gas.  Although it
appears from Figure~\ref{fig:hi} that this area is also devoid of
stars, it is coincidental that we did not obtain spectra there.
Optical imaging shows that stars do fill the H$\,${\sc i} hole.  The
contrast between structure in the gas and lack thereof in the stars
emphasizes that the gas traces the recent dynamical evolution of the
galaxy, whereas the stars have been relaxed over many dynamical times.

Aquarius is the only one of the three dIrrs to show rotation in the
gas.  The gas rotates about a north--south axis, with a velocity
differential of about 10~km~s$^{-1}$ from east to west.
Interestingly, the stellar distribution is elongated in the same
direction, as if the stars formed a disk that rotates in the same
plane as the gas.  However, the stellar velocities show no evidence
for rotation.  Therefore, the fact that the stellar isophotes and the
gas rotation share the same axis is either coincidental or reflective
of the underlying structure of dark matter.  For example, the dark
matter halo may be triaxial, with a long axis pointing east--west.
This is merely a speculative suggestion in the absence of many more
stellar velocity measurements that could constrain the velocity
anisotropy.  Additionally, spectra of the younger blue stars would
reveal to what degree only the old RGB stars are representative of the
entire stellar population.

SagDIG has an H$\,${\sc i} hole, like Leo~A\@.  It is possible that a
recent supernova explosion carved the hole in the gas.  The explosion
would not have affected the positions of the stars, which fill the
hole.  However, \citet{mom14} found that even young stars can be found
in the hole, which could indicate that the hole was formed by
gravitational instability rather than stellar feedback.  On the other
hand, a study of similar kpc-sized H$\,${\sc i} holes in five nearby
dIrrs found that the underlying stellar population produced sufficient
energy to create the holes \citep{war11}.  Ignoring the hole, the gas
density contours are fairly circular.  If the gas is rotating in a
disk, then we are viewing the disk face-on.  As a result, we would not
be able to detect any rotation along the line of sight.

\subsection{Comparison to Other Dwarf Galaxies}

\begin{figure}[t!]
\centering
\includegraphics[width=\columnwidth]{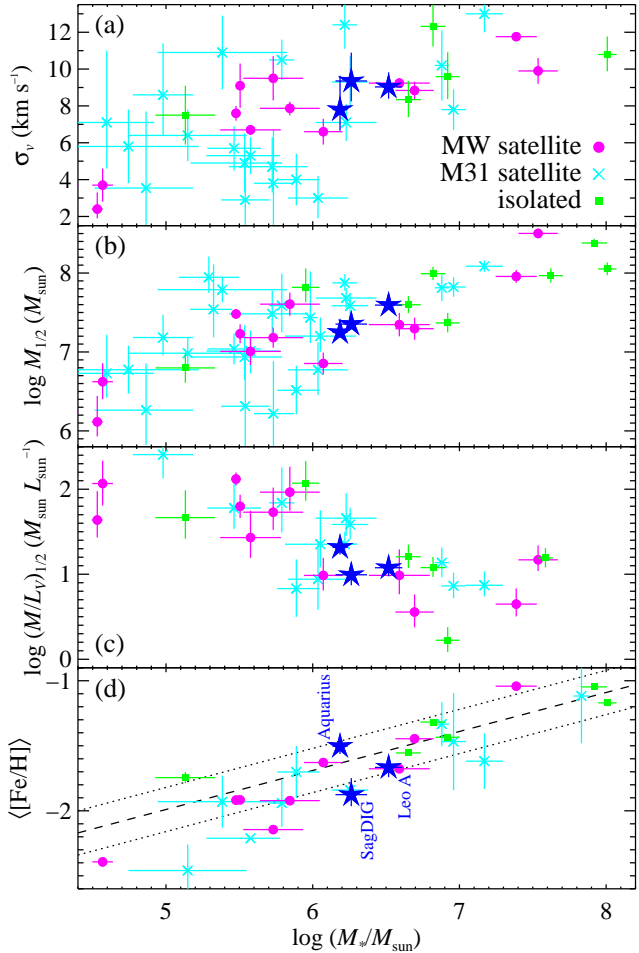}
\caption{The trend with stellar mass of (a) velocity dispersion, (b)
  mass within the half-light radius, (c) mass-to-light ratio within
  the half-light radius, and (d) average metallicity.  MW satellite
  galaxies are shown as magenta circles, and isolated galaxies in and
  around the Local Group are shown as green squares.  The dashed line
  in panel (d) shows the mass--metallicity relation from
  \citet{kir13b}.  The dotted lines show the 0.17~dex scatter about
  the relation.  The three isolated galaxies presented in this paper
  are shown as blue, five-pointed stars.  For other isolated galaxies,
  the dynamical quantities are taken from \citet[][Leo~T]{sim07},
  \citet[][Tucana]{fra09}, and \citet[][others]{kir14}.  For the MW
  satellites, the dynamical quantities are taken from \citet{sim07},
  \citet{mat08}, \citet{ade09}, \citet{koc09},
  \citet{wal07,walker09a,walker09b,walker09c}, \citet{kop11}, and
  \citet{fri12}.  The dynamical quantities for the M31 satellites are
  from \citet{tol12,tol13} and \citet{col13}.  All metallicities are
  from \citet{kir13b}.\label{fig:trends}}
\end{figure}

Figure~\ref{fig:trends} shows some dynamical properties of the three
dIrrs with other Local Group galaxies.  Velocity dispersion, dynamical
mass, and mass-to-light ratio within the half-light radius are shown
as a function of stellar mass.  These quantities follow well-defined
trends with stellar mass.  Velocity dispersion and mass increase and
mass-to-light ratio decreases with mass.  Leo~A, Aquarius, and SagDIG
fall well within the ranges of these quantities defined by other Local
Group galaxies.

The color coding in the plot distinguishes MW satellites from isolated
galaxies.  The three dIrrs in this paper fall in the latter group.
Regardless, the two groups do not show different trends.  The lack of
dichotomy in $\sigma_v$ and $M_{1/2}$ is not surprising.  The
environmental influence of the MW is not expected to drastically
affect the dark matter density in the center of the galaxy (the region
that dictates the stellar velocities) until just before the galaxy is
completely disrupted by tides \citep{pen08}.  Environment could
potentially affect $M/L_V$ because ram pressure stripping truncates
star formation.  The sudden cessation of stellar mass growth might be
expected to stunt galaxy growth, resulting in a lower luminosity for a
given mass.  Stripping also causes satellite galaxies to be older on
average than field galaxies, which would also reduce $L_V$ at a given
stellar mass.  Nonetheless, the satellite galaxies do not show a
higher $M/L_V$ on average than the isolated galaxies.  The effect of
environment on this quantity seems to be below our ability to detect
it.


\section{Chemical Composition}
\label{sec:chemistry}

The metallicity distributions of dwarf galaxies record the details of
the history of gas flow in the galaxies while stars were forming.  For
example, in a ``Closed Box'' model of chemical evolution
\citep{sch63,tal71}, the mean stellar metallicity approaches the
stellar yield as the gas supply runs out.  The model can be modified
to a ``Leaky Box'' by allowing gas to escape.  If the gas escapes at a
rate proportional to the SFR (${\rm outflow} = \eta \times {\rm
  SFR}$), then the yield, $p$, can simply be replaced by the effective
yield: $p_{\rm eff} = p/(1 + \eta)$.  In the limit where the true
yield is invariant among galaxies, then the average metallicity can be
used as a proxy for the amount of gas lost during the lifetime of star
formation.

Panel (d) of Figure~\ref{fig:trends} shows the relation between the
average stellar metallicities, $\langle {\rm [Fe/H]} \rangle$, and
stellar mass for dwarf galaxies.  Leo~A, Aquarius, and SagDIG fit in
the trend defined by other dwarf galaxies \citep{kir11a,kir13b}.  The
relation is linear for dwarf galaxies ($10^4 < M_*/M_{\sun} < 10^8$).
Larger galaxies continue this trend up to about $10^9~M_{\sun}$, and
even larger galaxies are continuous with the smaller galaxies, though
the slope flattens above $10^9~M_{\sun}$ \citep{gal05}.  Interpreted
in the context of the Leaky Box model, the larger galaxies have deeper
gravitational potential wells, which allows them to retain more gas
(smaller $\eta$) in the face of stellar feedback.

As \citet{kir13b} pointed out, dIrrs are indistinguishable from dSphs
on the mass--metallicity relation.  Despite still possessing gas and
forming stars, the average metallicity depends much more strongly on
stellar mass than on the amount of star formation the galaxy has yet
to complete.  However, the presence of gas allows an analysis not
possible for the dSphs: comparison between gas-phase and stellar
metallicities.\footnote{The gas-phase metallicity in Aquarius is not
  measurable because it has no current star formation and therefore no
  H$\,${\sc ii} regions from which gas-phase metallicity could be
  measured \citep{van97}.}  For example, Leo~A and SagDIG, as very low
luminosity dIrrs, have low metallicities appropriate for their low
stellar mass \citep{ber12}.  It turns out that they also lie low on
the stellar mass--stellar metallicity relation.

Their low metallicities could be a consequence of their high gas
fractions.  In the Closed Box model, the gas fraction decreases over
time, which establishes a monotonic, inverse proportionality between
gas fraction and metallicity.  \citet{pag97} presented the relations
between present metallicity, yield, gas fraction, and average stellar
metallicity.  The yield is $p = Z / (\ln \mu^{-1})$
(\citeauthor{pag97}'s Equation 8.6), where $Z$ is the current
gas-phase metallicity and $\mu$ is the gas fraction.  The gas
fractions for Leo~A and SagDIG are 0.69 and 0.82 (see
Table~\ref{tab:properties}).  The gas-phase metallicities are 4.9\%
\citep{vanzee06} and 4.0\% \citep{sav02} of the solar value.  The
yields, then, are $0.13~Z_{\sun}$ and $0.20~Z_{\sun}$.  The average
stellar metallicity is $p \left(1 + \frac{\mu \ln \mu}{1 -
  \mu}\right)$ (\citeauthor{pag97}'s Equation 8.8), corresponding to
${\rm [M/H]} = -1.6$ and $-1.7$ for Leo~A and SagDIG, respectively.
These values are strikingly close to our measurements of $\langle {\rm
  [Fe/H]} \rangle =
\leoafehmean_{-\leoafehmeanerrlower}^{+\leoafehmeanerrupper}$ and
$\sagdigfehmean_{-\sagdigfehmeanerrlower}^{+\sagdigfehmeanerrupper}$.

The preceding argument is very rough.  First, the conclusions depend
on how well the galaxy approximates a closed box.  A variable gas loss
rate could heavily affect the determination of the yield.  For
example, \citet{sav02} estimated that gas loss from SagDIG would lead
to a larger value for the ``true'' gas fraction.  Their prediction for
the stellar metallicity (${\rm [M/H]} = -2$) ended up being slightly
too low, but they were right to point out that gas loss needs to be
considered in a more careful model of chemical evolution.  Second, the
gas-phase metallicities are most sensitive to oxygen, whereas the
stellar metallicities are based on iron absorption lines.  Because the
O/Fe ratio is unknown, we implicitly assumed that the value is solar.
In reality, the value will depend on the changing ratio of Type~II and
Type~Ia supernova ejecta in the galaxy's chemical evolution.

\subsection{Metallicity Distributions}
\label{sec:mdf}

While the average metallicity of a galaxy does contain a lot of
information, resolved stellar spectroscopy affords an even more
valuable diagnostic of chemical evolution.  Specifically, the shape of
the metallicity distribution depends on the details of the gas flow
during star formation.  For example, the narrowness of the
distribution can indicate how much the galaxy violates the assumptions
of the Closed Box or Leaky Box.

\begin{figure*}[t!]
\centering
\includegraphics[width=0.9\textwidth]{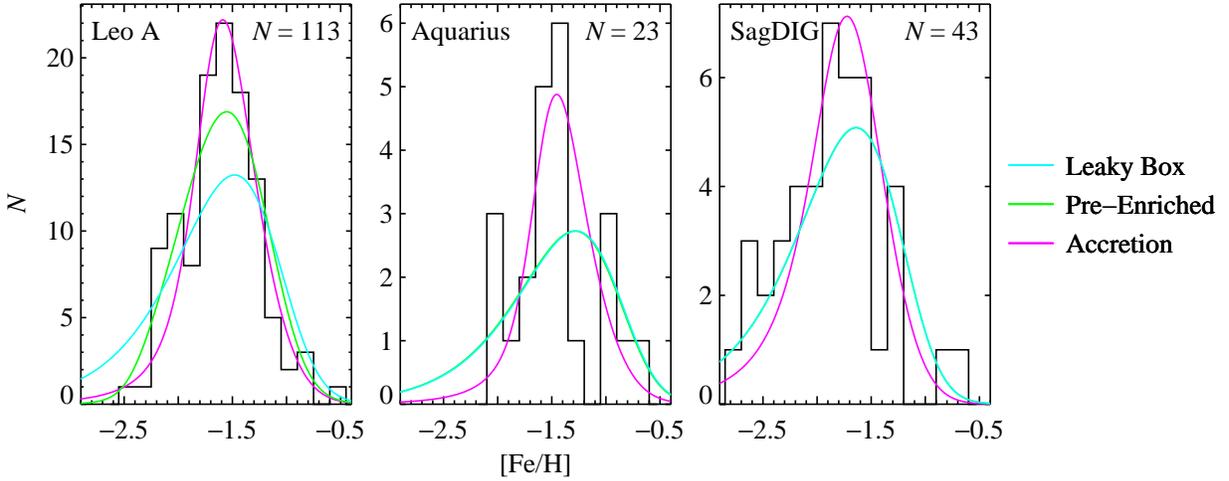}
\caption{Histograms of [Fe/H] in 0.15~dex bins.  The upper right
  corners show the number of stars in each histogram.  The colored
  curves show the best-fit chemical evolution
  models.\label{fig:fehhist}}
\end{figure*}

Figure~\ref{fig:fehhist} shows the metallicity distributions for the
three dIrrs.  Leo~A has the best sampled distribution, followed by
SagDIG and then Aquarius.  As a result, the quantitative results for
Leo~A will be the most secure.  Within the margins of uncertainty,
there is no significant difference between the shapes of the
metallicity distributions other than the mean metallicity.  The
skewness and kurtosis of all three distributions are consistent with
each other and with zero.  Interestingly, these shapes are more in
line with small dSphs, which had a limited star formation lifetime
\citep{kir13b}.  Leo~A, Aquarius, and SagDIG all had star formation
durations well in excess of several Gyr \citep{col07,col14,wei14a}.

Deductions concerning chemical evolution from the metallicity
distributions can be made more quantitative by using the equations of
chemical evolution.  Under certain assumptions, the equations predict
a stellar metallicity distribution.  The assumptions we invoke here
are homogeneity (a one-zone model), instantaneous mixing of gas, and
instantaneous recycling of stellar ejecta.  We invoke the same models
considered by \citet{kir11a,kir13b}.  Those models are the Leaky Box,
as discussed above; the Pre-Enriched model, which is the Leaky Box
that starts with non-zero metallicity; and the Accretion model, which
allows for gas inflow during star formation.

The Leaky Box predicts the following metallicity distribution:

\begin{equation}
\frac{dN}{d\mathfeh} \propto \left(\frac{10^{\mathfeh}}{p_{\rm eff}}\right) \exp \left(-\frac{10^{\mathfeh}}{p_{\rm eff}}\right) \: . \label{eq:leaky}
\end{equation}

\noindent
The only free parameter is $p_{\rm eff}$.  The Pre-Enriched Model
\citep{pag97} predicts a similar metallicity distribution:

\begin{equation}
\frac{dN}{d\mathfeh} \propto \left(\frac{10^{\mathfeh} - 10^{\mathfeh_0}}{p_{\rm eff}}\right) \exp \left(-\frac{10^{\mathfeh}}{p_{\rm eff}}\right) \: . \label{eq:preenriched}
\end{equation}

\noindent
The two free parameters are $p_{\rm eff}$ and $\mathfeh_0$, the
initial metallicity of the gas.  \citet{lyn75} invented the Accretion
model (called the Best Accretion model by \citeauthor{lyn75} and the
Extra Gas model by \citealt{kir11a}).  The infalling gas has zero
metallicity, and it falls in at a prescribed rate.  The metallicity
distribution is described by two transcendental equations that must be
solved for the stellar mass fraction, $s$.

\begin{eqnarray}
\nonumber \mathfeh(s) &=& \log \bigg\{p_{\rm eff} \left(\frac{M}{1 + s - \frac{s}{M}}\right)^2 \times \\
            & & \left[\ln \frac{1}{1 - \frac{s}{M}} - \frac{s}{M} \left(1 - \frac{1}{M}\right)\right]\bigg\}\label{eq:s} \\
\nonumber \frac{dN}{d\mathfeh} &\propto&  \frac{10^{\mathfeh}}{p_{\rm eff}} \times \\
            & & \frac{1 + s\left(1 - \frac{1}{M}\right)}{\left(1 - \frac{s}{M}\right)^{-1} - 2 \left(1 - \frac{1}{M}\right) \times 10^{\mathfeh}/p_{\rm eff}}\label{eq:infall}
\end{eqnarray}

\noindent
The two free parameters are $p_{\rm eff}$ and the accretion parameter,
$M$, which is the ratio of the final mass to the initial gas mass.

One major assumption of the models is that they assume the galaxy has
completed its star formation.  Otherwise, the prediction is only valid
for stars below a certain metallicity.  For the Leaky Box, that
metallicity is $p_{\rm eff} \ln \mu^{-1}$.  In fact, the model would
predict no stars to form above that metallicity, which is the present
gas-phase metallicity.  About half of the stars in the three dIrrs
have metallicities larger than this threshold.  Clearly, the dIrrs
violate some of the model assumptions.  Nonetheless, we fit these
models to examine how well their distributions conform to some of the
most basic predictions from chemical evolution.

We fit each of these models to each of the observed metallicity
distributions.  The likelihood that a galaxy's metallicity
distribution conforms to a model is

\begin{eqnarray}
\nonumber L &=& \prod_i \int_{-\infty}^{\infty} \frac{dP}{d\mathfeh} \frac{1}{\sqrt{2 \pi}\,\delta\mathfeh_i} \times \\
            & & \exp \left(-\frac{\left(\mathfeh - \mathfeh_i \right)^2}{2\left(\delta\mathfeh_i\right)^2}\right) d\mathfeh \label{eq:lprod}
\end{eqnarray}

\noindent
In Equation~\ref{eq:lprod}, $dP/d\mathfeh$ is the probability
distribution of the model, and $\mathfeh_i$ and $\delta\mathfeh_i$ are
the metallicity and error of the $i^{\rm th}$ observed star.  We used
an MCMC to maximize the likelihood that all of the measured
metallicities with errors less than 0.5~dex conform to the model.  The
length of the MCMC chains were $10^3$ for the Leaky Box and $10^5$ for
the other two models.  Table~\ref{tab:properties} gives the
best-fitting values and their 68\% confidence intervals.

We also computed the corrected Akaike information criterion
\citep[AICc,][]{aka74,sug78}, which quantifies how well one model fits
over another.  It penalizes models with more free parameters.  This is
especially important for the Pre-Enriched and Accretion models because
they are both generalizations of the Leaky Box model.  Therefore, it
is not possible for the Leaky Box model, which has only one free
parameter, to fit better than either of the other two models.  The
AICc allows for the Leaky Box model to be favored even if its formal
likelihood is lower.  The probability that one model fits better than
another is $\Delta (\ln P) = \Delta {\rm AICc}/2$.
Table~\ref{tab:properties} lists the values of $\Delta (\ln P)$ for
the Pre-Enriched and Accretion models as compared to the Leaky Box
model.  Negative values mean that the Leaky Box model is preferred
over the alternative model.

Formally, each galaxy prefers a different model.  Leo~A favors the
Pre-Enriched model.  Figure~\ref{fig:fehhist} shows that the
Pre-Enriched model does the best job at fitting the metal-poor end of
the distribution.  However, the Accretion model fits the peak better.
This ambivalence is reflected in the small difference between AICc for
the two models.  The Accretion model has a probability of
$\leoappreinfall$ of fitting the metallicity distribution of Leo~A
relative to the Pre-Enriched model.  Thus, neither model is strongly
preferred, but they are both strongly preferred relative to the Leaky
Box model.

Aquarius and SagDIG show no evidence for requiring any additional
complication beyond the Leaky Box.  Formally, Aquarius prefers the
Accretion model.  However, the difference of AICc between the
Accretion and Leaky Box models is negligible.  For both models, the
best fit value of ${\rm [Fe/H]}_0$ for the Pre-Enriched model was
extremely low.  As a result, the Pre-Enriched model is nearly
indistinguishable from the Leaky Box model.  Although the data do not
justify an additional free parameter beyond the Leaky Box, the
metallicity distributions are not as well sampled as Leo~A\@.  Our
result might be different with more measurements.

Again, the dIrrs violate some of the assumptions of these chemical
evolution models.  In particular, they have not run out of gas, so we
should not be fitting the distributions of the more metal-rich stars.
Furthermore, all three dIrrs are late-forming.  They must have
acquired gas or somehow made their gas available for star formation
long after their dark matter halos collapsed.  As a result, any gas
accretion that might have powered star formation almost certainly does
not follow the functional form described by Equation~\ref{eq:infall}.

\subsubsection{Correction for Selection Bias}
\label{sec:mdfcorr}

\begin{figure}[t!]
\centering
\includegraphics[width=\columnwidth]{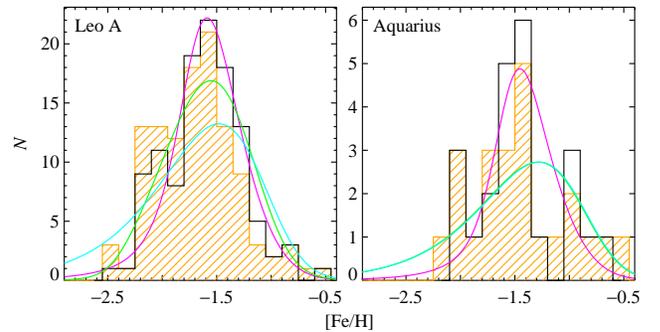}
\caption{Metallicity distributions for Leo~A and Aquarius.  The solid,
  black histograms are observed (same as Figure~\ref{fig:fehhist}).
  The shaded, orange histograms are corrected for RGB selection bias,
  as described in Section~\ref{sec:mdfcorr}.  The curves show the same
  chemical evolution models as
  Figure~\ref{fig:fehhist}.\label{fig:mdfcorr}}
\end{figure}

Red giant lifetimes are much less than a Hubble time, and they vary as
a function of stellar metallicity and mass.  As a result, the observed
distribution of red giant metallicities is not guaranteed to be an
exact representation of the actual mass fraction of heavy elements in
a stellar population that spans a wide range of ages.  Even in the
absence of any color selection, the varying red giant lifetimes bias
the observed metallicity distributions.  Manning \& Cole (in
preparation) are investigating the relationship between the
observational sample selection of red giant stars and the actual
underlying distribution of stellar metallicities.  They use synthetic
RGBs based on PARSEC isochrones \citep{bre12} to estimate the expected
number distribution of red giants, representing various populations in
the spectroscopic sample as a function of both age and metallicity,
allowing for scaling dictated by the SFH\@.  The net result is a
variation on the order of 20\% between the oldest, most metal-poor
stars and the youngest, most metal-rich stars.  Manning \& Cole will
present stellar number distributions scaled by the approximate over-
or under-representation of giants as a function of metallicity for
predominantly intermediate-age populations, like those in Leo~A and
Aquarius.
 
We calculated scaling factors for each bin in the metallicity
histograms (Figure~\ref{fig:fehhist}) based on the HST-derived SFHs
and age--metallicity relations of both Leo~A and Aquarius
\citep{col14}.  The shaded histograms in Figure~\ref{fig:mdfcorr} show
the expected number of stars for each metallicity bin of the entire
stellar population, whereas the unshaded histograms show the observed
number of red giants in each metallicity bin.  The SFHs were used to
estimate the mean population age, but the correction factors do not
depend strongly on the detailed variation of SFR with time.  We did
not compute corrections for SagDIG because the available photometry is
not as deep as for Leo~A or Aquarius.
 
With the corrections from Manning \& Cole (in preparation), we found
that the corrected mean metallicity of Leo~A is lower than the
observed distribution by 0.07~dex, whereas the mean metallicity of
Aquarius is unchanged.  The corrected histograms for both dIrrs are
slightly less concentrated toward the peak than the observed
histograms.

The changes in the mean and shape of the metallicity distribution will
affect the parameters derived for the chemical evolution models.  They
could also affect which model is preferred.  In the case of Leo~A, the
effective yields would be lower by an amount comparable to the shift
in mean metallicity.  The slightly less peaked distribution also would
cause the Pre-Enriched model to have an even higher likelihood
relative to the Accretion model.  However, the corrections still do
not allow for the Leaky Box model without pre-enrichment.  The
corrections in Aquarius are not severe enough to significantly affect
the parameters or likelihoods of the chemical evolution models.

Most dIrrs have not yet been subjected to scrutiny of the red giant
selection effect.  In order to compare galaxies on level footing, we
restrict our analysis to the uncorrected metallicity distributions.
None of the figures and tables except Figure~\ref{fig:mdfcorr} reflect
these corrections.

\subsection{Metallicity Gradients}

The metallicity of stars in a galaxy often decreases with distance
from the center of the galaxy.  A variety of scenarios could give rise
to a radially decreasing metallicity gradient.  For instance,
low-metallicity gas that falls onto the galaxy would likely have some
angular momentum.  That gas would end up in the outer regions of the
galaxy, diluting the metallicity of any gas that happened to be
forming stars.  Alternatively, if the galaxy loses gas, it will lose
it most readily from the outer parts of the galaxy, where the
gravitational potential is the weakest.  As a result, later star
formation is likely to come from gas surviving at the center of the
galaxy.

\begin{figure}[t!]
\centering
\includegraphics[width=0.90\columnwidth]{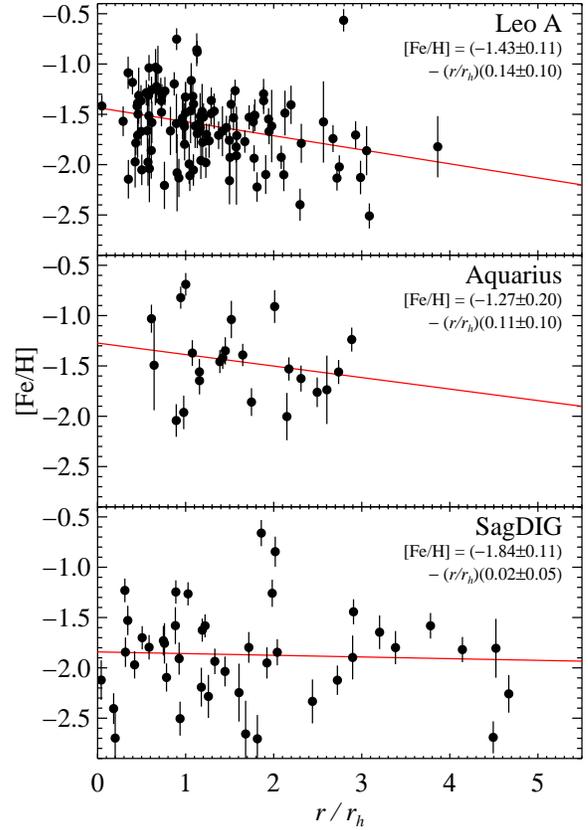}
\caption{The relation between metallicity and radial distance from the
  center of the galaxy in units of the half-light radius.  The red
  lines are least-squares linear regressions whose parameters are
  shown in the upper right corner of each panel.\label{fig:gradient}}
\end{figure}

Figure~\ref{fig:gradient} shows the radial metallicity gradients in
the three dIrrs.  All of the gradients are negative, in line with the
preceding arguments and consistent with most other dwarf galaxies.
However, the gradients are shallow.  The slopes in Leo~A and Aquarius
are only marginally distinct from zero, and the slope in SagDIG is
effectively zero.

A rigorous analysis of the metallicity gradient should sample the
metallicities of stars out to many half-light or core radii, ideally
out to the tidal radius.  Our survey samples much of the extent of
Aquarius and most of SagDIG\@.  However, Leo~A contains quite a few
stars beyond the extent of our spectroscopic sample.  Therefore, our
results are strictly applicable only to the range of stars we have
observed.  Some galaxies, especially dSphs, can have changes in the
slopes of their metallicity gradients that would not be detected in
stellar samples with limited angular extent \citep[e.g.,][]{bat11}.

\citet{lea13} found that the radial metallicity gradients in dIrrs are
usually shallower than dSphs.  Indeed, \citet{kir11a} found rather
steep gradients in some MW dSphs, though some of their spectroscopic
samples were limited to about 10~arcmin from the center of the dSph.
\citeauthor{lea13}\ further found that the gradients were the least
steep in galaxies with more rotational support.  For example, the
three highest luminosity dSph satellites of M31 \citep[NGC~147,
  NGC~185, and NGC~205; see][]{kor12} have metallicity gradients that
monotonically steepen as their rotational support ($v_{\rm
  rot}/\sigma_v$) decreases \citep{geh06,geh10,kol09}.  As
\citet{lea13}\ pointed out, rotation induces an angular momentum
barrier that prevents gas and consequently star formation from
concentrating in the center of the galaxy.  As a result, chemical
evolution and the increase of metallicity occur at similar rates at
all radii.  (See \citealt{schroyen11} for galaxy simulations that
support this hypothesis.)  Alternatively, radial migration could
flatten any pre-existing metallicity gradient.

We did not detect rotational support in the stellar populations of
Leo~A, Aquarius, or SagDIG (Section~\ref{sec:rot}).  However, our
observations only place limits on the amount of rotation.  They do not
conclusively rule out rotation.  This is especially true for SagDIG,
which shows a rather circular distribution of H$\,${\sc i} gas,
suggesting that we could be viewing the galaxy face-on.  Thus, our
observations neither support \citeauthor{lea13}'s finding that
metallicity gradients become shallow with increasing $v_{\rm
  rot}/\sigma_v$ nor provide evidence against it.  Larger samples
could help tighten the constraints on both $v_{\rm rot}/\sigma_v$ and
the metallicity gradients.

\subsection{The Age--Metallicity Relation}

The broadband colors of red giants depend on their ages and
metallicities.  Therefore, we can couple the photometry of the stars
with our spectroscopic measurements of their metallicities.  We used
Yonsei-Yale isochrones \citep{dem04} for this purpose.  First, we
isolated the two sets of isochrones that bordered a given metallicity.
We interpolated each isochrone in metallicity space to create a set of
isochrones at the exact metallicity of each star.  Then, we isolated
the two isochrones that bordered the $I_0$ magnitude and $(V-I)_0$
color of the star in question.  We linearly interpolated between the
two isochrones to estimate the age of that star.  For stars where we
measured $[\alpha/{\rm Fe}] \ge +0.3$, we used the $[\alpha/{\rm Fe}]
= +0.3$ isochrones.  For stars where we measured $[\alpha/{\rm Fe}]
\le 0.0$, we used the $[\alpha/{\rm Fe}] = 0.0$ isochrones.  For other
stars, we used linear interpolations of the two isochrone sets.

When we measured metallicities from the DEIMOS spectra, we used
photometry to fix the surface gravity and provide a best guess at
$T_{\rm eff}$.  In this process, we used Yonsei-Yale isochrones with
an assumed age of 14~Gyr.  Thus, it might seem circular to derive ages
from these metallicities.  However, the photometric temperature is
used only as a first guess.  The spectrum determines the final
temperature.  Furthermore, the color--temperature relations are not
particularly sensitive to age.  For example, consider a star with $M_V
= -2.3$ and $V-I=1.2$.  The photometric temperature for this star
would be 4400~K at 2~Gyr and 4477~K at 14~Gyr.  Even if we adopted
strictly photometric temperatures in the metallicity determinations,
the difference in [Fe/H] between those two temperatures would be less
than 0.1~dex \citep[e.g.,][]{kir10}.  In contrast, the metallicities
at those ages would be ${\rm [Fe/H]} = -1.29$ and $-1.91$,
respectively.  Hence, the spectroscopic metallicity does have the
power to determine ages, even if the spectroscopic metallicities were
based partly on photometric temperatures.

To estimate errors, we resampled the star's color, magnitude, and
metallicity.  For example, in one realization, we perturbed the
magnitude by $R\,\delta I_0$, where $R$ is a random number drawn from
a Gaussian distribution with unit width, and $\delta I_0$ is the
uncertainty on the magnitude.  The metallicity and color were
perturbed in a similar fashion, and we re-computed the age for this
combination of values.  We repeated this process $10^3$ times.  We
took the uncertainty on the age to be the standard deviation of all of
the trials.

\begin{figure}[t!]
\centering
\includegraphics[width=0.90\columnwidth]{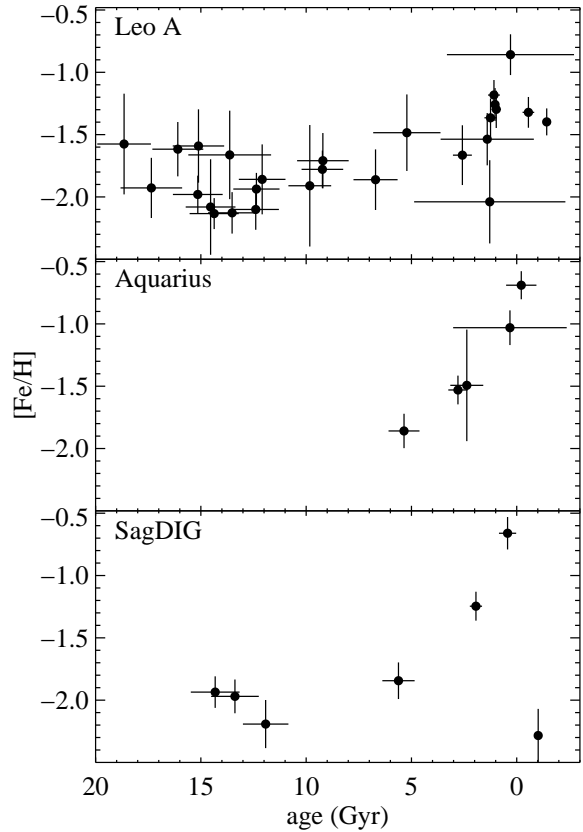}
\caption{The relation between metallicity and age for individual
  stars.  The red lines are orthogonal linear regressions whose
  parameters are shown in the upper left corner of each
  panel.\label{fig:agefeh}}
\end{figure}

Figure~\ref{fig:agefeh} shows metallicities as a function of the
resulting ages.  Measurements with uncertainties larger than 4~Gyr are
excluded.  Some ages are younger than zero or older than the age of
the Universe.  These measurements reflect measurement uncertainty or
imperfect isochrone models.  Even so, the metallicity mostly increases
with time in all cases, as expected for most galaxies.

\citet{col07} derived a very shallow age--metallicity relation for
Leo~A\@.  They found that a model with ${\rm [Fe/H]} = -1.4$, constant
with time, fit the CMD well.  We also found a shallow age--metallicity
relation, but it turns up within the last 5~Gyr.

\citet{col14} also measured the photometric age--metallicity relation
for Aquarius.  They found a positive slope up to 6~Gyr ago, followed
by virtually no metallicity evolution thereafter.  In contrast, we
found a steep age--metallicity relation.  However, the oldest age we
measured is between 5 and 6~Gyr.  We attribute the difference in
results to two factors.  First, we used ground-based photometry,
whereas \citeauthor{col14}\ used HST photometry, which is more
accurate.  Second, \citeauthor{col14}\ fit entire stellar populations,
whereas we fit individual stars.  Third, we have access to the
spectroscopic metallicities.  In principle, this information should
allow a fairly precise measurement of age, but in practice, random
errors in the photometry and systematic errors in the isochrones limit
the precision of the measurement.

\subsection{$\alpha$ Elements}

Chemical abundance ratios provide a method of estimating SFHs
complementary to CMD fitting.  The trend of the [$\alpha$/Fe] ratio
with [Fe/H] is sensitive to the changing ratio of Type~II to Type~Ia
supernovae with time.  A numerical chemical evolution model---as
opposed to the analytic models in Section~\ref{sec:mdf}---can predict
the trend of [$\alpha$/Fe] versus [Fe/H] for an assumed SFH
\citep[e.g.,][]{mat86,kir11b}.

\begin{figure}[t!]
\centering
\includegraphics[width=\columnwidth]{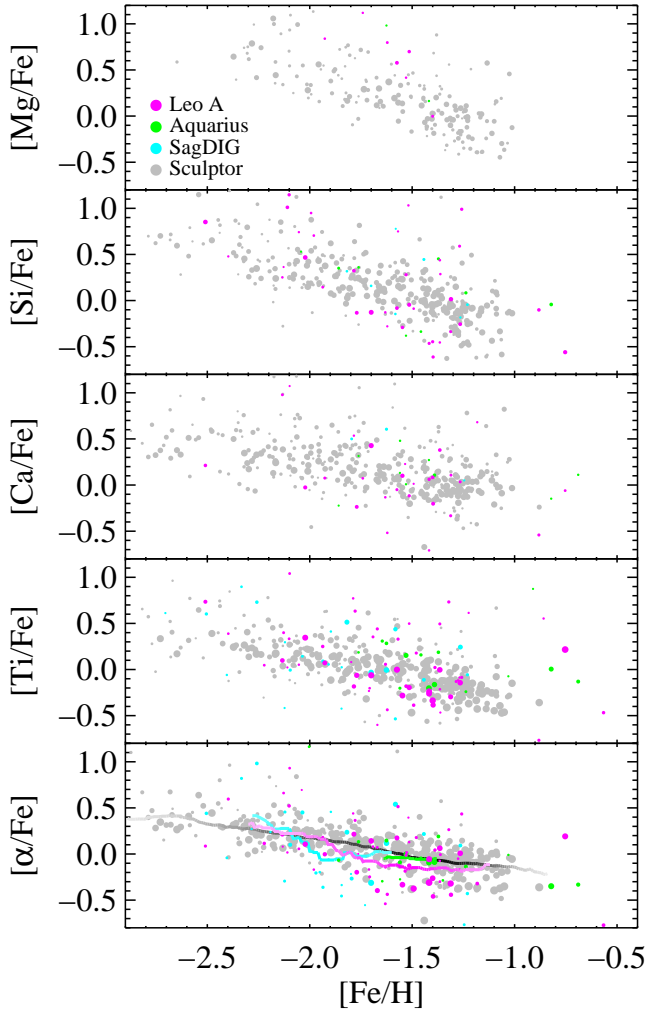}
\caption{The trend of abundance ratios with metallicity.  The three
  dIrrs are shown in color.  For comparison, the Sculptor dSph is
  shown in gray \citep[data from][]{kir09,kir10}.  Point size is
  inversely proportional to measurement uncertainty, where the largest
  points have uncertainties of about 0.1~dex in each dimension, and
  the smallest points (barely visible) have uncertainties of about
  0.5~dex.  The bottom panel shows the average [$\alpha$/Fe] ratio,
  including moving averages in window of $\Delta{\rm [Fe/H]} = 0.5$.
  The weight of the lines representing the moving averages reflects
  the number of stars contributing to the average at each
  [Fe/H].\label{fig:alpha}}
\end{figure}

Figure~\ref{fig:alpha} shows the trend of [$\alpha$/Fe] with [Fe/H]
for the three dIrrs.  The MW satellite Sculptor, a dSph, is shown for
comparison.  The scarcity of points and the measurement uncertainties
limit our ability to draw strong conclusions about the chemical
evolution of the dIrrs.  For example, fitting chemical evolution
models in the style of \citet{kir11a} would not be informative.
Regardless, it is worth pointing out that this is by far the largest
sample of [$\alpha$/Fe] measurements in dIrrs.\footnote{However, it is
  not the largest sample outside of the MW system.  See
  \citet{var14a,var14b} for [$\alpha$/Fe] measurements in M31 and its
  satellites.}

However, the similarity of the dIrrs to Sculptor is interesting,
especially in the bottom panel, which shows the average of the
$\alpha$ elements.  For the most part, the dIrr measurements fall
within the envelope defined by Sculptor.  Some of the highly uncertain
measurements scatter upward to $[\alpha/{\rm Fe}] > +0.5$.  These
measurements are simply uncertain, and they probably do not
necessarily reflect the presence of extremely $\alpha$-enhanced stars.

However, some of the more secure measurements in the dIrrs lie on the
low side or even entirely below the Sculptor distribution.  The moving
averages for the dIrrs---especially Leo~A---lie slightly below that of
Sculptor at ${\rm [Fe/H]} > -1.9$.  The underabundance of $\alpha$
elements is most easily interpreted as a SFH that is more extended
than in Sculptor.  The longer time allows more Type~Ia supernovae to
explode, lowering the [$\alpha$/Fe] ratio.  Reassuringly, the
photometrically measured SFHs for the dIrrs are significantly more
extended than Sculptor.  Whereas star formation persisted for 8~Gyr or
more in the dIrrs, Sculptor formed no stars after 10~Gyr ago
\citep{rev09,wei14a}.


\section{Summary and Discussion}
\label{sec:summary}

We have presented a comprehensive spectroscopic analysis of the
stellar populations in the three dIrrs Leo~A, Aquarius, and SagDIG\@.
These galaxies share several properties.  First, they have similar
stellar masses, spanning a range of only 1.5--$3.3 \times
10^6~M_{\sun}$.  Second, they are isolated.  The dwarf galaxies
nearest to them are hundreds of kpc away, and the nearest large
galaxies (the MW, M31, and M33) are at least 600~kpc away.  Second,
they formed the majority of their stars well into the lifetime of the
Universe.  Deep HST CMDs show that Leo~A and Aquarius formed 80--90\%
of their stars more recently than 6 and 8~Gyr ago, respectively
\citep{col07,col14}.  Although SagDIG does not yet have HST imaging
that reaches the MSTO, the CMD of the more luminous stars indicates a
similar SFH but with a possibly larger fraction of ancient stars
\citep{wei14a}.

Their similarities make it even more interesting to study their subtle
differences.  For example, dwarf galaxies obey a very tight relation
between stellar mass and stellar metallicity \citep{kir11a,kir13b}.
Both gas-rich dIrrs and gas-poor dSphs all fall on the same relation,
with no measurable difference between them.  The three dIrrs we have
studied fall within 1.6 standard deviations of the relation.  However,
the two most gas-rich galaxies, Leo~A and SagDIG, fall below the line,
whereas Aquarius lies above it.  This result is consistent with the
expectations of simple models of chemical evolution.  As galaxies
evolve, they become metal-rich and gas-poor.  Hence, gas-rich galaxies
have not yet become metal-rich.

The dynamical properties of the three dIrrs are in line with other
dwarf galaxies.  Their velocity dispersions, masses, and mass-to-light
ratios follow the same trend as other Local Group galaxies.
Table~\ref{tab:properties} shows these properties.  One quantity of
particular interest is the ratio of total mass to the baryonic mass,
including stars and H$\,${\sc i} gas.  Aquarius, has the highest
ratio.  It is easy to imagine that Aquarius appears similar to future
versions of Leo~A and SagDIG\@.  If the latter two galaxies continue
to turn gas into stars, their gas masses and SFRs will decrease.
Their dark-to-baryonic mass ratios will increase along with their
gas-phase and average stellar metallicities.  The final states of
Leo~A and SagDIG will contain more stellar mass than Aquarius because
they have more gas available for star formation.  Other than that,
they seem to be less evolved---but slightly more massive---versions of
Aquarius.

The observed RGB stars do not show any sign of rotation in any of the
three dIrrs.  As \citet{whe15} concluded, dIrrs do not seem to
transition to dSphs by eliminating well-ordered rotation.  Instead,
the lowest mass dwarf galaxies form dynamically hot, supported by
dispersion rather than rotation.  It is notable that the H$\,${\sc i}
gas in Aquarius does rotate.  In fact, the structure and motions of
gas in all of the galaxies has little relation to the old stellar
populations.  Whereas the gas is clumpy and exhibits small-scale
velocity structure, the stellar distribution is smooth.  The only
similarities between the gas and the old stars are the average
velocities and the velocity dispersions.  The similarity of velocity
dispersions supports the conclusion that the gas is reacting to the
galaxies' gravitational potentials rather than hydrodynamics.

The gold standard of measuring SFHs is space-based photometry deep
enough to reach the old MSTO\@.  Nonetheless, we have employed
complementary techniques to evaluate the ages of stars.  First, we fit
the ages of stars to model isochrones based on their spectroscopic
metallicities.  Although the details of the age--metallicity relation
do not match the photometrically derived SFHs, we found a monotonic
slope in the relation such that younger stars are more metal-rich.
Second, we qualitatively described the SFH as compared to the Sculptor
dSph, a satellite of the MW, by comparing the [$\alpha$/Fe]
distributions.  On average, the dIrr stars have lower [$\alpha$/Fe] at
a given [Fe/H] than Sculptor.  This distinction indicates more of an
influence of Type~Ia supernovae in the dIrrs than in Sculptor.  This
result is in agreement with the photometric SFHs because longer star
formation durations allow for more Type~Ia supernovae.

It is also worthwhile to compare these dIrrs with dIrrs of different
masses.  Two dIrrs with detailed studies are WLM \citep{lea12,lea13}
and NGC~6822 \citep{swa16}, which have stellar masses of $4.5 \times
10^7~M_{\sun}$ and $1.7 \times 10^8~M_{\sun}$ \citep{woo08}, both
considerably more massive than the dIrrs studied in our work.
\citet{lea12} presented a rich dynamical data set for WLM\@.  They
clearly detected stellar rotation, even though $v_{\rm rot}/\sigma_v
\sim 1$.\footnote{\citet{whe15} found a lower value for $v_{\rm
    rot}/\sigma_v$, but they also found that WLM exhibited clear
  stellar rotation.  Furthermore, \citet{lea12} corrected for
  asymmetric drift, which led to a larger value of $v_{\rm rot}$.}
The gas has a similar rotation curve but a velocity dispersion smaller
by a factor of six.  Therefore, WLM seems to have experienced
dynamical evolution markedly different from the three dIrrs in our
sample.  \citeauthor{lea12}\ suggested that giant molecular clouds
inflated the velocity dispersion of the stars over time.
\citeauthor{lea13} found that the metallicity distribution of WLM,
like Leo~A, does not conform to a Leaky Box.  \citeauthor{swa16}'s
(\citeyear{swa16}) spectroscopic study of NGC~6822 tells a similar
story.  Over time, the stars became dynamically heated, such that the
younger stars have lower velocity dispersions than the older stars.
Similar to Aquarius, the gas in NGC~6822 rotates whereas the stars
show only scant evidence for rotation \citep[also see][]{kir14}.  Like
WLM, the younger, more metal-rich stars can be found closer to the
center.

Our study, as well as other studies of Local Group dIrrs, paints a
picture of galaxies tenuously forming stars.  Although Leo~A,
Aquarius, and SagDIG first started forming stars near the beginning of
the Universe, they managed to form very few stars for billions of
years.  They experienced low, simmering SFRs since they resumed star
formation.  Small events could disrupt the gas distribution of the
entire galaxy.  At any given time, the gas appears in disordered
clumps, giving rise to star formation sporadic in both time and
position.

Despite teetering at the threshold of the ability to form stars, the
galaxies managed to evolve chemically for many Gyr.  Their increase in
metallicity was slow (0.05 to 0.30~dex~Gyr$^{-1}$), but they managed
to maintain that pace for up to 8~Gyr.  Furthermore, they follow clear
dynamical and chemical trends with stellar mass.  The predictable
outcome of these galaxies despite such a tenuous existence suggests
that they are perhaps not so fragile after all.

\acknowledgments

We thank Alan McConnachie, Nobuo Arimoto, and Mike Irwin for kindly
sharing their photometry catalog for Aquarius.  We also thank Yazan
Momany and his collaborators for sharing their photometry catalog of
SagDIG\@.  Deidre Hunter provided kind assistance with the LITTLE
THINGS 21~cm images.  We acknowledge support from the National Science
Foundation through grant 1614081.

{\it Facility:} \facility{Keck:II (DEIMOS)}

\bibliography{dirr2016}
\bibliographystyle{apj}

\end{document}